\documentclass[journal]{IEEEtran}

\usepackage{amsmath,amssymb,amsthm}
\usepackage{accents}  %

\usepackage{graphicx}
\usepackage[dvipsnames]{xcolor}

\usepackage{hyperref}
\usepackage{cite}
\usepackage{balance}

\usepackage[capitalise]{cleveref}
\crefname{equation}{}{}  %
\crefname{figure}{Fig.}{Figs.}

\usepackage{pgfplots}
\pgfplotsset{compat=1.16,
            line1/.style = {line width=1.0pt},
            marker1/.style = {mark options={scale=1.4, line width=0.4pt, fill=white}}, 
            MDSR/.style = {color=Fuchsia, line1, mark=square*, marker1},
            MDSRD/.style = {color=RoyalBlue, line1, mark=*, marker1},
            LTR/.style = {color=Peach, line1, mark=triangle*, marker1},
            Total/.style = {color=OliveGreen, line1, mark=triangle*, marker1},
            converse/.style = {dashed, line width=0.5pt},
            PG/.style = {color=Green, line1, mark=pentagon*, marker1},
            BD-IR/.style = {color=OliveGreen, line1, mark=diamond*, marker1},
}
\tikzset{
            nodetext/.style = {anchor=west, font=\scriptsize},
}
\usetikzlibrary{shapes, arrows}

\newcommand{\bs}[1]{\boldsymbol{#1}}
\newcommand{\msf}[1]{\mathsf{#1}}

\newcommand{\ceil}[1]{\lceil#1\rceil}
\newcommand{\floor}[1]{\lfloor#1\rfloor}
\newcommand{\E}{\mathbb{E}}  %
\newcommand{\T}{\mathsf{T}}  %

\newcommand{\GFq}{\text{GF}(q)}
\newcommand{\GF}[1]{\text{GF}(#1)}

\newcommand{\ro}{R_\msf{o}}
\newcommand{\rt}{R_\msf{i}}

\newcommand{\ciTX}{\bs c_i^\T\bs X}
\newcommand{\ciTx}{\bs c_i^\T\bs x}
\newcommand{\rowsofW}{\bs w_1^\T,\ldots,\bs w_k^\T}
\newcommand{\rowsofC}{\bs c_1^\T,\ldots,\bs c_{k/\ro}^\T}
\newcommand{\uncodedprods}{\bs w_1^\T\bs X,\ldots,\bs w_k^\T\bs X}
\newcommand{\codedprods}{\bs c_1^\T\bs X,\ldots,\bs c_{k/\ro}^\T\bs X}

\newcommand{\ncu}{n_\msf{u}}
\newcommand{\nce}{n_\msf{e}}
\newcommand{\fcpu}{f_\msf{cpu}}

\newcommand{\Lambold}{\bs\Lambda}
\newcommand{\lambold}{\bs\lambda}

\newcommand{\Lc}{L_\mathsf{c}}

\newcommand{\Ld}{L_\mathsf{d}}
\newcommand{\Ldu}{L_\mathsf{d}^\msf{u}}
\newcommand{\ldu}{l_\mathsf{d}^\msf{u}}
\newcommand{\Lde}{L_\mathsf{d}^\msf{e}}

\newcommand{\Ldec}{L_\msf{dec}}
\newcommand{\Ldecu}{L_\msf{dec}^\msf{u}}
\newcommand{\Ldece}{L_\msf{dec}^\msf{e}}

\newcommand{\tauu}{\tau_\msf{u}}
\newcommand{\taue}{\tau_\msf{e}}

\newcommand{\Na}{N_\msf{a}}
\newcommand{\Nm}{N_\msf{m}}
\newcommand{\Nau}{N_\msf{a}^\msf{u}}
\newcommand{\Nmu}{N_\msf{m}^\msf{u}}
\newcommand{\Nae}{N_\msf{a}^\msf{e}}
\newcommand{\Nme}{N_\msf{m}^\msf{e}}

\newcommand{\thickbar}{\rule{.4em}{.8pt}}
\newcommand{\thickunderbar}[1]{\underaccent{\thickbar}{#1}}
\newcommand{\Lclower}{\thickunderbar{L}_\msf{c}}
\newcommand{\Ldulower}{\thickunderbar{L}_\msf{d}^\msf{u}}
\newcommand{\ldulower}{\thickunderbar{l}_\msf{d}^\msf{u}}

\newcommand{\carda}{\left|\mathcal{A}\right|} %
\newcommand{\cardb}{\left|\mathcal{B}\right|}%

\newcommand{\calv}{\mathcal{V}}
\newcommand{\cala}{\mathcal{A}}
\newcommand{\calb}{\mathcal{B}}
\newcommand{\cals}{\mathcal{S}}
\newcommand{\calc}{\mathcal{C}}

\newcommand{\ceilm}{\lceil\bar{m}\rceil}
\newcommand{\floorm}{\lfloor\bar{m}\rfloor}

\newcommand{\Pf}{\text{P}_\msf{f}}
\newcommand{\PF}{\text{P}_\msf{F}}
\newcommand{\cfac}{\frac{u\log_2(q)}{\nu}}

\newtheorem{theorem}{Theorem}

\newtheorem{lemma}{Lemma}
\newtheorem{example}{Example}

\newcommand{\bx}{\boldsymbol{x}}
\newcommand{\bX}{\boldsymbol{X}}
\newcommand{\bW}{\boldsymbol{W}}

\begin{document}

\title{Rateless Codes for Low-Latency Distributed Inference in Mobile Edge Computing}

\author{Anton~Frigård,
        Siddhartha~Kumar,
        Eirik~Rosnes,~\IEEEmembership{Senior~Member,~IEEE},\\
        and Alexandre~Graell~i~Amat,~\IEEEmembership{Senior~Member,~IEEE}%
\thanks{This work was financially supported by the Swedish Research Council under grant 2020-03687. This paper will be presented in part at the 17th International Symposium on Wireless Communication Systems (ISWCS), Berlin, Germany, Sep. 2021 \cite{Fri21}.}%
\thanks{A. Frigård, S. Kumar, and E. Rosnes are with Simula UiB, Bergen, Norway, e-mail: \{anton,~kumarsi,~eirikrosnes\}@simula.no.}%
\thanks{A. Graell i Amat is with the Department of Electrical Engineering, Chalmers University of Technology, Gothenburg, Sweden, e-mail: alexandre.graell@chalmers.se, and with Simula UiB, Bergen, Norway.}
}
\maketitle

\begin{abstract}
We consider a mobile edge computing scenario where users  want to perform a linear inference operation $\bs W\bs x$ on local data $\bx$ for some network-side matrix $\bs W$. The inference is performed in a distributed fashion over multiple servers at the network edge. For this scenario, we propose a coding scheme that combines a rateless code to provide resiliency against straggling servers\textemdash hence reducing the computation latency\textemdash and an irregular-repetition code to  provide spatial diversity\textemdash hence reducing the communication latency. We further derive a lower bound on the total latency\textemdash comprising  computation latency, communication latency, and decoding latency. The proposed scheme performs remarkably close to the bound and yields significantly lower latency than the scheme based on maximum distance separable codes recently proposed by Zhang and Simeone.

\end{abstract}

\IEEEpeerreviewmaketitle

\section{Introduction}

Performing heavy processing on resource-constrained devices may result in unacceptably high latency and energy consumption. This is often dealt with by outsourcing computation tasks to the \textit{cloud}, i.e., remote data centers. This strategy is inadequate for latency-critical applications, however, since it suffers from long propagation delays and congestion in the core and backhaul networks \cite{Mach2017}. Mobile edge computing (MEC) is a concept aimed at eliminating these shortcomings by providing computing capacity closer to where data is generated \cite{Mach2017,ETSI2015}. In MEC, users offload  computations to servers located at the network edge, which process the data and return the results.

An MEC system functions like a distributed computing system and therefore shares some of its afflictions, such as stragglers, i.e., servers that take an unacceptably long time to finish their tasks \cite{Dean2013}. The straggler problem
has been addressed in the neighboring field of distributed computing in data centers by means of coding. By treating stragglers as erasures, redundant computations can be introduced by applying erasure correcting codes, such that the overall computation task can be completed from the subtasks from a subset of the servers (via a decoding step) \cite{Ananthanarayanan,Wang2015,Lee2018,Yu2017,Sev19,Mallick2019,Sev2019Raptor}. Replication of subtasks\textemdash i.e., using a trivial erasure correcting code\textemdash was investigated in \cite{Ananthanarayanan,Wang2015}. Maximum distance separable (MDS) codes were considered in  \cite{Lee2018} for distributed gradient descent. In \cite{Yu2017}, a coding strategy referred to as polynomial codes was proposed for large-scale matrix-matrix multiplication and showed to minimize the number of servers to wait for. Most  works on
coded distributed computing disregard the impact of decoding on the total latency. An exception is 
 \cite{Sev19,Sev2019Raptor}, where it was shown that the conventional application of MDS codes suffers from high decoding latency, which may severely impact the total latency.

Whereas research on resource-allocation in MEC is extensive \cite{Sard2015,Zhang2016,Dinh2017,Chen2018,Wang2016,Wang2017,Zhang2018,Li2020}, only  few works have considered coding in the context of MEC \cite{Zhang2019,Li2021,Kumar2021_arxiv}. 
In \cite{Zhang2019}, the authors considered  the scenario in which multiple devices want to perform a linear inference  $\bW\bx$ on some local data $\bx$ given a network-side model matrix $\bW$. Such operations arise in, e.g., collaborative filtering recommender systems \cite{Schafer2007}.  Particularly, the authors in \cite{Zhang2019} proposed a coding scheme that combines an MDS code to provide resiliency against stragglers, and (regular) replication of computations across edge servers, whose primary goal  is to create spatial diversity such that  edge servers can send the results back to multiple users simultaneously using cooperative zero-forcing precoding, exploiting the nature of the wireless channel \cite{Zhang2019Fund}. Thus, a trade-off between computation and communication latency is induced, whereby more MDS coding provides more straggler mitigation at the expense of  less repetition and, therefore, less spatial diversity\textemdash hence higher communication latency\textemdash while more replication has the opposite effect. The same coding scheme was used in \cite{Li2021} for the linear inference problem $\bW\bX$, where $\bs X$ denotes a matrix. The scheme in \cite{Zhang2019,Li2021} requires a subset of the edge servers to complete their assigned subtasks, i.e., it does not exploit partial computations that some straggling servers might have performed. Exploiting partial results from servers was proposed in  \cite{Sev19,Sev2019Raptor} in the context of coded distributed computing by using a rateless code. Furthermore, as most works on coded distributed computing, \cite{Zhang2019,Li2021} ignore the impact of the decoding latency on the total latency. In \cite{Kumar2021_arxiv}, we proposed a block-diagonal scheme based on partitioning $\bW$ into smaller submatrices and applying MDS codes to each submatrix, which also exploits partial results. This scheme outperforms the scheme in \cite{Zhang2019}, where most of the gain comes from the block-diagonal structure, which allows to reduce the decoding latency.

In this paper, we consider the same distributed inference problem $\bW\bx$ as in 
\cite{Zhang2019}. For this scenario,  we propose a scheme that
combines a rateless code\textemdash more precisely a Luby-transform (LT) code\textemdash for straggler mitigation
and an irregular-repetition code that replicates computations
across edge servers to create spatial diversity. The rateless
code leads to a scheme that naturally exploits partial results. Compared to
the regular-repetition code considered in \cite{Zhang2019,Li2021}, the proposed irregular-repetition
code enables additional degrees of freedom in the overall code design. We notice   that exploiting partial results by the servers can also be applied to a coding scheme with an underlying MDS code. We therefore propose a scheme that augments the scheme in \cite{Zhang2019,Li2021} by exploiting partial results and utilizing  irregular repetition, yielding more flexibility in the code design  and the assignment of computations to the servers. %
The proposed scheme based on rateless codes yields significantly lower total latency\textemdash comprising computation latency, communication
latency, and decoding latency\textemdash than the scheme in \cite{Zhang2019} and  better performance than our scheme in \cite{Kumar2021_arxiv}. Moreover, unlike \cite{Zhang2019,Li2021,Kumar2021_arxiv}, where decoding is performed at the user devices\textemdash albeit its cost on the total latency is ignored in \cite{Zhang2019,Li2021}\textemdash we also let the edge servers perform the decoding step, which leads to a further reduction in the total decoding latency. The proposed scheme based on MDS codes and irregular repetition also outperforms the scheme in \cite{Zhang2019}, but performs poorer than the scheme based on rateless codes. 
We finally derive a converse (lower) bound on the total latency. Remarkably, the proposed scheme based on rateless codes performs very close to the bound for the considered scenarios.

\emph{Notation}:
We will denote matrices by bold, uppercase letters, vectors by bold, lowercase letters, sets by calligraphic, uppercase letters, and random variables by uppercase letters, e.g., \(\bs X\), \(\bs x\), \(\mathcal{X}\), and \(X\) represent a matrix, a column vector, a set, and a random variable, respectively. We also use uppercase letters for some variables; their (non-random) nature can be easily inferred from the context.
The transpose of a vector or a matrix  is denoted as $(\cdot)^\msf{T}$. Random vectors will, like matrices, be denoted by bold, uppercase letters. Their stochastic nature will however be explicitly stated to avoid confusion. For some positive integer $a$, we denote by $[a]$ the set $\{1,\ldots,a\}$. %
For some real number $b$, the notation $(b)^+$ is defined as $\max\{b, 0\}$. A finite field of order $q$ will be denoted by $\GFq$. The set of natural numbers is denoted by $\mathbb{N}$. The expectation of a random variable is denoted by $\E[\cdot]$.

\section{System Model}
We consider a wireless network consisting of $u$ single-antenna user devices and \(e\) single-antenna edge servers\textemdash hereafter referred to as edge nodes (ENs)\textemdash where we assume that $e\leq u$, for simplicity. Each user $j$ has some length-$r$ data vector $\bs x_j$ and wants to compute the linear inference operation $\bs y_j = \bs W \bs x_j$, for some network-side $k\times r$ matrix $\bs W$ over $\GFq$. By denoting the rows of $\bs W$ as $\rowsofW$ and forming the matrix $\bs X = [\bs x_1,\ldots,\bs x_u]$ of size $r\times u$, the goal is equivalent to computing $\uncodedprods$. We assume that $\bs W$ is static and up-to-date for a sufficiently long period of time such that it can be stored by the ENs offline. 

The ENs are prone to straggling and $\bs W$ is therefore encoded for protection prior to being stored by the ENs. Specifically, each length-$k$ column of
$\bs W$ is encoded by an \([n_1, k]\) erasure correcting code of code rate $\ro=k/n_1$ to form a coded matrix $\bs C$ of size $\frac{k}{\ro} \times r$.\footnote{Technically, the code is defined over a subfield of $\GFq$. As a general remark, all codes considered in this paper are defined over a subfield of the data field $\GF{q}$ even if this is not be explicitly mentioned. The actual subfield considered for the numerical results is specified in \cref{sec:results}.} We refer to the code as the \textit{straggler code} and to the rows of $\bs C$, denoted by $\rowsofC$, as the coded rows. Each coded row is replicated $1/\rt$ times on average across ENs, where $\rt=n_1/n$ is the  rate of replication and $n\geq n_1$ is the total number of replicated coded rows. The replication can be seen as an \([n, n_1]\) irregular-repetition code of code rate $\rt$. The combination of the straggler code and the irregular-repetition code can then be seen as a concatenated code of dimension $k$ and length $n=k/(\ro\rt)$ over a subfield of $\GF{q}$. The code is used to recover the desired results $\uncodedprods$ from the fraction of $\codedprods$ that were computed by the quickest ENs, by using a decoding algorithm. We hereafter refer to a product $\ciTX$ simply as a product.

\begin{figure}
    \centering
    \makeatletter
    \if@twocolumn
        \includegraphics[width=1.0\columnwidth]{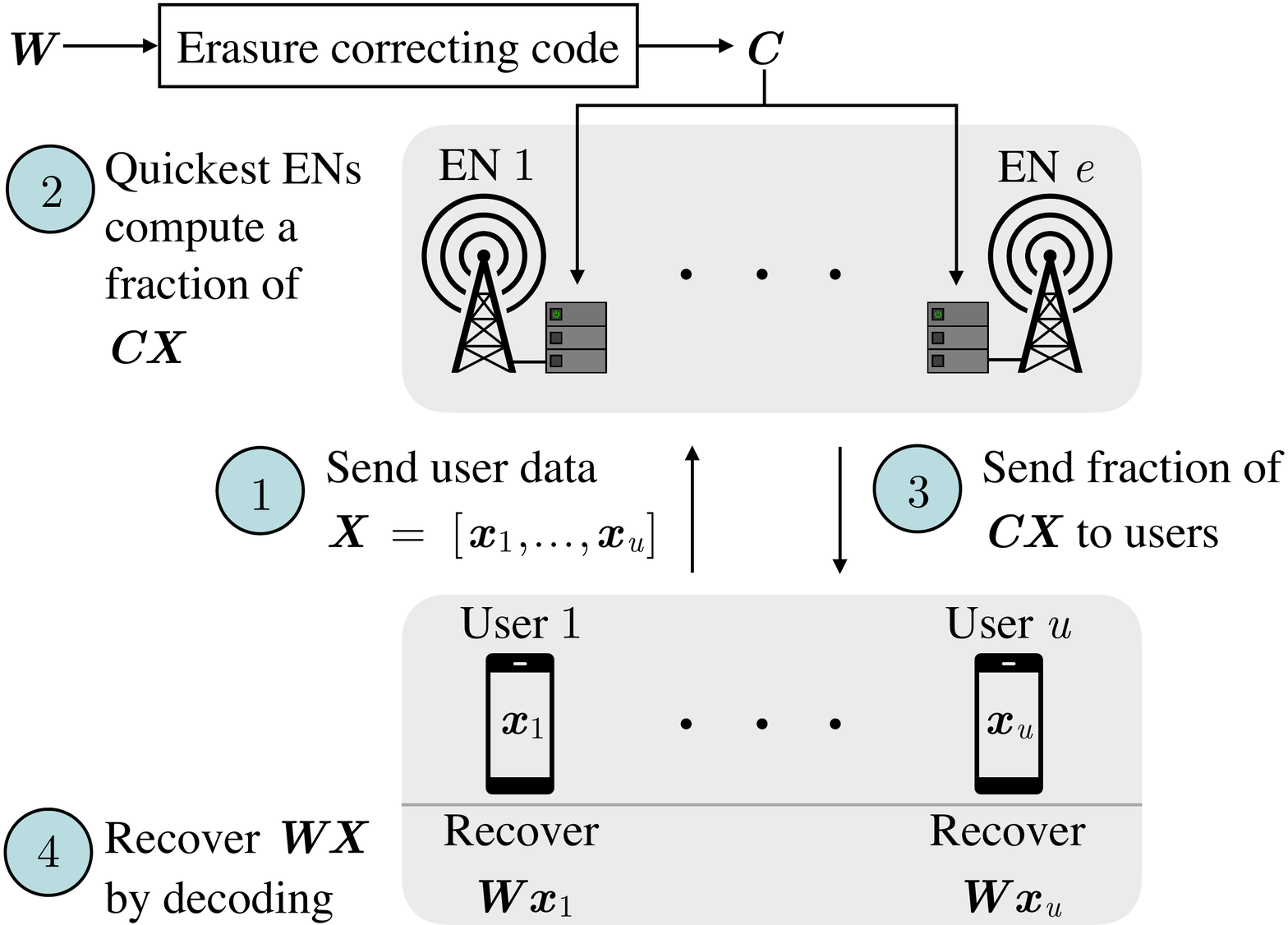}
    \else
        \includegraphics[width=0.5\columnwidth]{Figures/MEC_network.pdf}
    \fi
    \makeatother
    \caption{MEC network with $u$ users and \(e\) ENs. The offloading consists of the uplink, downlink, computation, and decoding phases.}
    \label{fig:system_model}
\end{figure}

Each EN thus stores $n/e=k/(\ro\rt e)$ coded rows. %
We define by $\mu$ the maximum number of coded rows that each EN can store, normalized by $k$. Thus,
\begin{align}
    \frac{k}{\ro\rt e} \leq \mu k \label{eq:storage_constraint}.
\end{align}

The coded rows assigned to each EN are placed in a queue determining the order in which they will be multiplied by $\bs X$. The queues are described by an \textit{assignment matrix} $\bs A$ of size $\frac{k}{\ro\rt e} \times e$, where entry $a_{ij}$ is the index of the $i$th coded row in the queue of EN $j$. As an example, if the third coded row in the queue of EN 2 is $\bs c_4^\T$, then $a_{32}=4$.

The offloading process consists of four phases\textemdash uplink, computation, downlink, and decoding\textemdash which are depicted in \Cref{fig:system_model}. We explain each phase in more detail next.

\subsection{Uplink Phase}
Initially, the users transmit their data to the ENs such that each EN can construct $\bs X$. We assume that the transmission scheme is fixed and always yields the same transmission latency.

\subsection{Computation Phase}
As soon as each EN has constructed $\bs X$, the computation phase begins, marked by time $t=0$. We model the straggling of EN $j$ as a time period of length $\Lambda_j$ at the start of the computation during which the EN is busy executing tasks from other applications \cite{Dean2013}. Following previous work \cite{Mallick2019, Zhang2019}, we let $\Lambda_1,\ldots,\Lambda_e$ be independent and identically distributed (i.i.d.) exponential random variables with mean $\beta$. We refer to $\Lambda_1, \ldots,\Lambda_e$ as the \textit{straggling times} and to $\beta$ as the \textit{straggling parameter}. We also define the vector $\Lambold = [\Lambda_1,\ldots,\Lambda_e]$.

After straggling, EN $j$ computes products $\bs c_{a_{ij}}^\T\bs X$, for $i\in[k/(\ro\rt e)]$. For simplicity, we assume a homogeneous setup where all ENs have the same number of CPU cores $\nce$ operating at clock frequency $\fcpu$. %
We then model the time it takes for one EN to compute an addition or a multiplication over $\GFq$ as $1/(\nce\fcpu)$ seconds (s).  Define $\delta$ as the time it takes for an EN to compute a product. %
Since this requires $u(r-1)$ additions and $ur$ multiplications, we have $\delta = (u(r-1) + ur)/(\nce \fcpu)$. Further, let $\bs D(t) = [D_1(t),\ldots,D_e(t)]$, where
\begin{align}
    D_j(t) = \min\bigg\{\bigg\lfloor\frac{(t-\Lambda_j)^+}{\delta}\bigg\rfloor, \frac{k}{\ro\rt e} \bigg\} \label{eq:Di_def}
\end{align}
is the number of products completed by EN $j$ at time $t$. We say that $\bs D(t)$ is a \emph{stopping vector} if we can recover the desired results $\uncodedprods$ by decoding the computed products at time $t$. To control the computation phase, we define a \textit{stopping set} $\cals$ of stopping vectors such that the computation phase ends as soon as $\bs D(t)\in\cals$. The computation latency is therefore
\begin{align}
     \Lc = \min \{t:\ \bs D(t)\in\cals\}. \label{eq:Lc_def}
\end{align}
By designing $\mathcal{S}$, we thus have some control over $\Lc$. This can be used to trade off computation latency for communication latency, as we will discuss in \Cref{sec:rateless_IR,sec:MDS_IR}. Furthermore, we note that the total number of products $P$ computed by the ENs at time $\Lc$ is given by
\begin{align}
    P = \sum_{j=1}^e D_j(\Lc). \label{eq:P_def}
\end{align}

Since a coded row $\bs c_i^\T$ may be stored by multiple ENs, the product $\ciTX$ may have been computed more than once. We denote by $M_i$, $i\in[k/\ro]$, the number of times product $\ciTX$ has been computed at the end of the computation phase, where $0 \leq M_i \leq e$. We will hereafter refer to $M_i$ as the spatial diversity of product $\bs c_i^\T \bs X$.

After the computation phase, we can recover the desired results $\uncodedprods$ by decoding the fraction of $\codedprods$ that were computed. In \cite{Zhang2019,Li2021}, the users are assumed to carry out the decoding, but the contribution of the decoding latency on the total latency is ignored. Since the idea of offloading is for users to save energy and to leverage the superior processing capabilities of the ENs, we also consider the scenario where the ENs decode. In the following two subsections, we outline each of these decoding configurations, which will be compared in \cref{sec:results}.

\subsection{Decoding by the Users}
\subsubsection{Downlink Phase}
We adopt the same interference channel model and transmission strategy as in \cite{Zhang2019}, which is based on \cite{Zhang2019Fund}. The signal received by each user is a weighted linear combination of the signals from all ENs plus Gaussian noise. The ENs are assumed to have full knowledge of the channel weights.

Consider specifically the transmission of product $\ciTX$, which is a length-$u$ vector. Let $J_i$ be the least common multiple of $M_i$ and $u$, $J_i=\mathrm{lcm}(M_i,u)$. We first divide each $\log_2(q)$-bit element of the product into $J_i/u$ packets of size $u\log_2(q)/J_i$ bits. The transmission is divided into $J_i/M_i$ blocks and in each block, $M_i$ distinct packets are transmitted in parallel to $M_i$ distinct users using zero-forcing precoding \cite{Zhang2019Fund}. The time to transmit the product is thus $u\log_2(q)/(M_i \nu)$ seconds, where $\nu$ (bits/s) is the transmission rate of an EN. At the end of this transmission, user $j$ has recovered $\ciTx_j$. 

The products are transmitted in sequence, and the communication latency is therefore
\begin{align}
    \Ldu = \frac{u\log_2(q)}{\nu} \sum_{\substack{i=1\\ M_i\geq 1} }^{k/\ro} \frac{1}{M_i}. \label{eq:Ldu_def}
\end{align}
When all products have been transmitted, user $j$ has recovered $\bs C'\bs x_j$, where $\bs C'$ is the matrix of coded rows that correspond to the products that have been computed. As an example, if product $\ciTX$ has been computed, then $\bs c_i^\msf{T}$ is a row in $\bs C'$.

Note that the packet size $u\log_2(q)/J_i$ is not necessarily an integer. Since we cannot transmit fractions of bits, one way is to pad each $\log_2(q)$ bit element of the product with extra bits such that it can be evenly divided into $J_i/u$ packets. In this case, the communication latency would be higher than what is given in \eqref{eq:Ldu_def}. However, numerical evaluations show that the communication latency is low enough such that the error is insignificant.

\subsubsection{Decoding Phase}
For simplicity, we assume that decoding starts as soon as each user $j$ has received $\bs C'\bs x_j$ and that each user is equipped with $\ncu$ CPU cores, operating at clock frequency $\fcpu$. Denote by $\Nau$ the number of additions and by $\Nmu$ the number of multiplications required to decode. The decoding latency is then
\begin{align}
    \Ldecu = \frac{1}{\ncu \fcpu} (\Nau + \Nmu). \label{eq:Ldecu_def}
\end{align}

\subsection{Decoding by the Edge Nodes}
\subsubsection{Decoding Phase}
Only the ENs that have stopped straggling partake in the decoding. These ENs communicate their computed products to each other via a high-speed backhaul link, such that each EN stores all computed products and can recover $\uncodedprods$ using a decoding algorithm. We assume that the decoding is prioritized by the ENs, such that no EN straggles by executing tasks from other applications. The decoding latency is given by 
\begin{align}
     \Ldec^\msf{e} = \frac{1}{n_\msf{e} \fcpu}(\Nae + \Nme), \label{eq:Ldece_def}
\end{align}
where $\Nae$ and $\Nme$ denote the number of additions and multiplications, respectively, that are needed for an EN to decode $u$ vectors.

\subsubsection{Downlink Phase}
The transmission of the recovered products $\uncodedprods$ works as follows. Denote by $M$ the number of ENs that are not straggling at time $\Lc$ and define $J=\mathrm{lcm}(M,u)$. Consider the transmission of the length-$u$ vector $\bs w_i^\T\bs X$ over $\GFq$. Divide each element of the vector into $J/u$ packets of size $u\log_2(q)/J$ bits and the transmission into $J/M$ blocks. In each block, $M$ distinct packets are transmitted to $M$ distinct users using zero-forcing precoding \cite{Zhang2019Fund}. The time to transmit one packet is $u\log_2(q)/(M\nu)$ seconds. 

By transmitting all products $\uncodedprods$ in sequence, the communication latency becomes
\begin{align}
     \Ld^\msf{e} = \frac{uk\log_2(q)}{\nu M}. \label{eq:Lde_def}
\end{align}
At the end of the transmission, user $j$ has recovered $\bs y_j=\bs W\bs x_j$.

\subsection{Problem Formulation} \label{subsec:problem_form}
The performance of the system is measured in terms of the total latency. If the users decode, the total latency is given by
\begin{align}
    \tauu = \E[ \Lc] + \E[ \Ldecu] + \E[ \Ldu], \label{eq:tauu_def}
\end{align}
whereas if the ENs decode, the total latency is given by
\begin{align}
    \taue = \E[ \Lc] + \E[ \Ldece] + \E[ \Lde]. \label{eq:taue_def}
\end{align}
The expectation is over $\Lambold$. Since the uplink transmission latency is the same for any coding scheme, it is not included in the total latency. Note that our freedom of design encompasses the code, assignment matrix $\bs A$, and stopping set $\cals$. We will therefore specifically refer to this combination as the \textit{coding scheme} and denote it by $\calc$. The overarching goal is to design a coding scheme that minimizes $\tauu$ or $\taue$.

\section{Preliminaries}
This section provides some basics on LT and MDS codes, since these codes are the base of our proposed coding schemes. We also describe the main scheme from \cite{Zhang2019}, which will be used as a benchmark.

\subsection{Luby-Transform Codes} \label{subsec:LT_code}
As a special class of rateless codes, LT codes \cite{Luby} are the first to perform close to capacity on erasure channels. Code symbols are formed as random linear combinations of the input symbols, and can be generated continuously until the receiver successfully decodes. Thus, the code rate does not have to be fixed beforehand and the code is therefore said to be rateless.

To generate a code symbol, a degree $d$ is drawn at random from a degree distribution $\Omega$. Then, $d$ out of the $k$ information symbols are chosen uniformly at random without replacement. For each such symbol, a weight from a subfield of $\GFq$ (the subfield chosen for the code) is chosen uniformly at random, and the sum of weighted symbols forms the code symbol. The so-called robust Soliton distribution \cite{Luby} is commonly used as the degree distribution. This distribution is determined by two parameters: i) the location of the spike, $\gamma$, which in the original paper \cite{Luby} is denoted by $k/R$; and ii) the parameter $\zeta$, which in \cite{Luby} is denoted by $\delta$ and is an upper bound on the probability of decoding failure, which is discussed next.

The number of code symbols that the receiver needs to collect in order to successfully decode is $k+\Phi$, where $\Phi$ is random. If we decide to fix an overhead $\phi$ and collect $k+\phi$ code symbols, there is therefore a nonzero probability of decoding failure, which we define as $\PF(\phi)\triangleq \text{Pr}(\Phi > \phi)$. Using peeling decoding, it was shown in \cite{Luby} that $\Phi\to 0$ as $k\to\infty$.

In this paper, we use inactivation decoding \cite{inactivation_dec_patent,shokrollahi2006raptor,Lazaro2017}, a maximum likelihood decoding algorithm which can be seen as a modification of peeling decoding. While peeling decoding might fail even if the received coded symbols can be successfully decoded (as it is not a maximum likelihood decoding algorithm), the inactivation decoder guarantees successful decoding if such is the case. This decreases the probability of decoding failure at the cost of a slight increase in complexity. In \cite{Schotsch2013}, it was shown that $\PF(\phi)$ for $q$-ary LT codes under maximum likelihood decoding can be upper-bounded as 
\makeatletter
\if@twocolumn
    \begin{align}
        &\PF(\phi) \leq \nonumber\\ &\sum_{i=1}^{k}\binom{k}{i}(q-1)^{i-1}
        \left(\frac{1}{q}+\frac{q-1}{q} \sum_{d} \Omega_{d} \frac{h_{d}(i ; k)}{h_{d}(0 ; k)}\right)^{k+\phi}, \label{eq:ML_upbound}
    \end{align}
\else
    \begin{align}
        \PF(\phi) \leq \sum_{i=1}^{k}\binom{k}{i}(q-1)^{i-1}
        \left(\frac{1}{q}+\frac{q-1}{q} \sum_{d} \Omega_{d} \frac{h_{d}(i ; k)}{h_{d}(0 ; k)}\right)^{k+\phi}, \label{eq:ML_upbound}
    \end{align}
\fi
\makeatother
where $k+\phi$ is the number of received symbols, $\Omega_d$ is the probability of generating degree $d$, and $h_\zeta(\xi; \sigma)$ is the Krawtchouk polynomial
\begin{align*}
    h_\zeta(\xi; \sigma) = \sum_{j=0}^\zeta (-1)^j (q-1)^{\zeta-j}\binom{\xi}{j} \binom{\sigma-\xi}{\zeta-j}.
\end{align*}
The bound is tight for all values of $\phi$ and considerably less computationally demanding than computing the true decoding failure probability $\PF(\phi)$.

\subsection{Reed-Solomon Codes} \label{subsec:rs_codes}
Reed-Solomon codes are a class of MDS codes. Specifically, an $[n_1,k]$ RS code of  rate $\ro$ can be successfully decoded using any set of $k$ coded symbols. In this paper, we use the Berlekamp-Massey (BM) decoding algorithm. The main parts of the decoding are the computation of the discrete Fourier transform of the received symbol sequence, and the feeding of this into a linear feedback shift register (LFSR) circuit \cite{Garrammone2013}. We compute the transform using the split-radix fast Fourier transform (FFT) \cite{Yavne1968}. The length of the input vector to the FFT must be a power of two. To this end, let $\eta = \ceil{\log_2(k/\ro)}$. If $k/\ro$ is not a power of two, the sequence is zero-padded to length $2^\eta$. The FFT is then computed using $2^{\eta-1}(3\eta-5)+4$ additions and $2^{\eta-1}(\eta-3)+2$ multiplications over $\GFq$ \cite{Yavne1968}.

For the LFSR step, let $F$ be the fraction of coded symbols that have been erased. The number of additions and multiplications (over $\GF{q}$) in this step are thus $\frac{k}{\ro}\big(\frac{k}{\ro} F-1\big)$ and $\big(\frac{k}{\ro}\big)^2 F$, respectively \cite{Garrammone2013}. The number of operations for BM decoding of a length-$k/\ro$ sequence can then be modeled as
\begin{align}
    \Na(k, \ro, F) &= 2^{\eta-1}(3\eta-5) + 4 + \bigg( \frac{k}{\ro}\bigg)^2 F - \frac{k}{\ro}, \label{eq:MDS_Na} \\
    \Nm(k, \ro, F) &= 2^{\eta-1}(\eta-3) + 2 + \bigg( \frac{k}{\ro}\bigg)^2 F. \label{eq:MDS_Nm}
\end{align}

\subsection{MDS-Repetition Scheme} \label{subsec:zhangs}
We here describe the  scheme from \cite{Zhang2019,Li2021}, which will be compared to our proposed schemes in \cref{sec:results}. The scheme uses an MDS code and regular replication of the MDS-coded rows to provide both straggler protection and spatial diversity. Henceforth, we will refer to this scheme as the MDS-repetition (MDS-R) scheme. Specifically, the columns of $\bs W$ are first encoded using an \([n_1, k]\) MDS code of code rate $\ro=k/n_1$, resulting in a matrix with coded rows $\rowsofC$. Each coded row is then stored by $1/\rt$ ENs, where $\rt=n_1/n$ is the rate of replication and $n\geq n_1$, restricted according to \eqref{eq:storage_constraint}, is the total number of replicated coded rows. Note that $1/\rt$ must be an integer. The assignment of coded rows is done by first splitting the $k/\ro$ MDS-coded rows into $\binom{e}{1/\rt}$ batches and then assigning each batch to $1/\rt$ ENs in a combinatorial fashion \cite{Zhang2019improved}.

\begin{example}
For $e=5$, $k=15$, $\ro=3/4$, and $\rt=1/3$, we have $20$ coded rows, divided into $10$ batches $\{\bs c^\T_1,\bs c^\T_2\}, \{\bs c^\T_3,\bs c^\T_4\},\ldots, \{\bs c^\T_{19},\bs c^\T_{20}\}$. Each batch is then assigned to $1/\rt = 3$ ENs, which are chosen as one out of $\binom{e}{1/\rt} = \binom{5}{3}$ possible combinations. The assignment matrix is thus
\makeatletter
\if@twocolumn
\begin{align*}
    \left(\begin{array}{ccccc}
        1 & 1 & 1 & 3 & 5 \\
        2 & 2 & 2 & 4 & 6 \\
        3 & 3 & 7 & 7 & 9 \\
        4 & 4 & 8 & 8 & {10} \\
        5 & 5 & 9 & {11} & {11} \\
        6 & 6 & {10} & {12} & {12} \\
        7 & {13} & {13} & {13} & {15} \\
        8 & {14} & {14} & {14} & {16} \\
        9 & {15} & {15} & {17} & {17} \\
        {10} & {16} & {16} & {18} & {18} \\
        {11} & {17} & {19} & {19} & {19} \\
        {12} & {18} & {20} & {20} & {20}    
    \end{array}\right).
\end{align*}
\else
\begin{align*}
\small{
    \left(\begin{array}{ccccc}
        1 & 1 & 1 & 3 & 5 \\
        2 & 2 & 2 & 4 & 6 \\
        3 & 3 & 7 & 7 & 9 \\
        4 & 4 & 8 & 8 & {10} \\
        5 & 5 & 9 & {11} & {11} \\
        6 & 6 & {10} & {12} & {12} \\
        7 & {13} & {13} & {13} & {15} \\
        8 & {14} & {14} & {14} & {16} \\
        9 & {15} & {15} & {17} & {17} \\
        {10} & {16} & {16} & {18} & {18} \\
        {11} & {17} & {19} & {19} & {19} \\
        {12} & {18} & {20} & {20} & {20}    
    \end{array}\right)}.
\end{align*}
\fi
\makeatother
\end{example}

The stopping set is designed such that the desired results $\uncodedprods$ can be recovered when $\xi$ ENs have completed all of their assigned products. The coding scheme is parameterized by the triplet $(\xi, \ro, \rt)$ and it can be shown that these parameters need to satisfy
\(\binom{e}{1/\rt} - \binom{e-\xi}{1/\ro} \geq \ro\binom{e}{1/\rt}\)
for decoding to be successful \cite{Zhang2019improved_arxiv}. 

In \cite{Zhang2019,Li2021}, the decoding latency was not taken into account. We therefore extend this scheme in \Cref{app:zhangs} by designing a decoding step. We also include an analysis of the total latency.

\section{Rateless Irregular-Repetition Scheme}
\label{sec:rateless_IR}
We propose a scheme that uses an LT code as straggler code, coupled with irregular replication of the LT-coded rows. We will refer to this scheme as the rateless irregular-repetition (Rateless-IR) scheme. As opposed to the MDS-R scheme \cite{Zhang2019,Li2021} outlined in \Cref{subsec:zhangs}, the Rateless-IR scheme collects partial results, i.e., it also collects products from ENs that have only completed a fraction of their assignment. The irregularity of the replication and the utilization of partial results are aimed at yielding more freedom in the system design than that provided by the MDS-R scheme. Additionally, the assignment matrix is not the same as for the MDS-R scheme.

Specifically, the columns of $\bs W$ are first encoded by an \([n_1, k]\) LT code %
of rate $\ro=k/n_1$, resulting in a matrix of rows $\rowsofC$. Recall from \cref{subsec:LT_code} that $k+\Phi$ coded symbols (distinct products in our case) are needed for successful decoding, where $\Phi$ is a random overhead. Since $\Phi$ cannot realistically be determined by the ENs during runtime, we choose instead to collect a fixed number of distinct products $k+\phi'$ and to design the degree distribution $\Omega$ such that the probability of decoding failure does not exceed a given target probability $\Pf$ at the overhead $\phi'$, i.e., $\PF(\phi') \leq \Pf$. The LT code is decoded using inactivation decoding \cite{inactivation_dec_patent,shokrollahi2006raptor,Lazaro2017}. We also choose the robust Soliton distribution \cite{Luby} as the degree distribution. For $q=2$, $\phi'= 2\mu k$  gives a large enough overhead such  that  the  decoding  latency  is  low,  while  also  keeping  a sufficiently  low  computation and communication latency.  We optimize the parameters $\gamma$ and $\zeta$ from \Cref{subsec:LT_code}, subject to the constraint $\PF(\phi') \leq \Pf$, in order to minimize the total number of decoding operations. The failure probability is approximated by \eqref{eq:ML_upbound}.

Each of the LT-coded rows is then replicated \textit{on average} $1/\rt$ times, where $\rt=n_1/n$ is the rate of replication and $n\geq n_1$ is the total number of replicated coded rows. Note that, as opposed to the MDS-R scheme \cite{Zhang2019,Li2021}, $1/\rt$ is not necessarily an integer. Furthermore, recall that $\ro$ and $\rt$ are constrained by \eqref{eq:storage_constraint}. By rearranging \eqref{eq:storage_constraint}, we see that $\ro \geq \frac{k}{\rt e\mu k} \geq \frac{k}{e\mu k}$, where the last inequality is a consequence of $\rt \leq 1$. Thus,
\begin{align*}
    \ro &\in \bigg\{1, \frac{k}{k+\phi'}, \frac{k}{k+\phi'+1}, \ldots, \frac{k}{e\mu k}\bigg\},
\end{align*}
where $\ro=1$ corresponds to using pure repetition, i.e., no LT code. Rearranging \eqref{eq:storage_constraint} again, we see that $\rt \geq \frac{1}{e\mu \ro}$. Therefore,
\begin{align*}
    \rt &\in\bigg\{1,\frac{k/\ro}{k/\ro + 1}, \ldots, \frac{k/\ro}{e\mu k}\bigg\}.
\end{align*}
In addition, we require that $k/(e\ro)$ (and $k/(e\ro\rt)$) are integers to make sure that no row in the assignment matrix is partly filled.

The assignment matrix is formed as follows. First, a matrix $\bs A_1$ of size $\frac{k}{e\ro} \times e$ is formed by assigning the numbers $1, 2,\ldots, k/\ro$ in ascending order from left to right, top to bottom. Next, a matrix $\bs A_2$ is formed by circularly shifting the rows of $\bs A_1$ one step to the left. This procedure is performed $\floor{1/\rt}$ times in total, where each new matrix $\bs A_i$ is formed by a single left-circular shift of the rows of the previous matrix $\bs A_{i-1}$. If $1/\rt$ is not an integer, a final matrix $\bs A_{\ceil{1/\rt}}$ is formed by circularly shifting the $(1 - \ceil{1/\rt} + 1/\rt)\frac{k}{e\ro}$ first rows of $\bs A_{\ceil{1/\rt}-1}$ one step to the left. The assignment matrix is then $\bs A = [\bs A_1^\T \cdots \bs A_{\ceil{1/\rt}}^\T]^\T$.

\begin{example}
For $e=5$, $k=15$, $\ro=3/4$, and $\rt=2/5$, the assignment matrix is
\makeatletter
\if@twocolumn
\begin{align*}
    \left(\begin{array}{ccccc}
        1 & 2 & 3 & 4 & 5 \\
        6 & 7 & 8 & 9 & 10 \\
        11 & 12 & 13 & 14 & 15 \\
        16 & 17 & 18 & 19 & 20 \\
        \hline
        2 & 3 & 4 & 5 & 1 \\
        7 & 8 & 9 & 10 & 6 \\
        12 & 13 & 14 & 15 & 11 \\
        17 & 18 & 19 & 20 & 16 \\
        \hline
        3 & 4 & 5 & 1 & 2 \\
        8 & 9 & 10 & 6 & 7 \\
    \end{array}\right),
\end{align*}
\else
\begin{align*}
\small{
    \left(\begin{array}{ccccc}
        1 & 2 & 3 & 4 & 5 \\
        6 & 7 & 8 & 9 & 10 \\
        11 & 12 & 13 & 14 & 15 \\
        16 & 17 & 18 & 19 & 20 \\
        \hline
        2 & 3 & 4 & 5 & 1 \\
        7 & 8 & 9 & 10 & 6 \\
        12 & 13 & 14 & 15 & 11 \\
        17 & 18 & 19 & 20 & 16 \\
        \hline
        3 & 4 & 5 & 1 & 2 \\
        8 & 9 & 10 & 6 & 7 \\
    \end{array}\right)},
\end{align*}
\fi
\makeatother
where the horizontal lines are used to highlight the component submatrices.
\end{example}

The goal is then for the ENs to compute a large enough fraction of $\codedprods$ such that $\uncodedprods$ can be recovered with high probability. While the results could be recovered with probability at least $1-\Pf$ if $k+\phi'$ distinct products are computed, we note that the total latency may in fact decrease by computing more products. This is because the spatial diversity of some products may decrease, which would decrease the communication latency in \eqref{eq:Ldu_def}, if the users decode, or the communication latency in \eqref{eq:Lde_def} (by increasing $M$), if the ENs decode. This will, however, increase the computation latency.

To find the optimal trade-off, we introduce a parameter $p$ and define the stopping set as the set of stopping vectors that guarantee the computation of $k+\phi'$ distinct products and \textit{at least} $p$ products in total. For simplicity, we also choose to discard products other than the $k+\phi'$ distinct products with highest spatial diversity. Note that, if the users decode, this will remove the summands in \eqref{eq:Ldu_def} corresponding to the discarded products, which will decrease the communication latency. On the contrary, if the ENs decode, it may decrease $M$ and therefore increase the communication latency in \eqref{eq:Lde_def}. Finally, the rows of $\bs C'$ corresponding to the discarded products are also removed.

Let us now consider decoding by the users and decoding by the ENs separately.

\subsection{Latency Analysis}
\subsubsection*{Decoding by the Users}
The communication latency is given by \eqref{eq:Ldu_def}. The number of decoding operations $\Nau$ and $\Nmu$ can be simulated, giving the decoding latency in \eqref{eq:Ldecu_def}. The total latency is then given by
\begin{align}
    \tauu = \min_{p, \ro, \rt} \bigg(\E[\Lc] + \frac{1}{\ncu \fcpu}\E[\Nau + \Nmu] + \E[\Ldu]\bigg), \label{eq:rateless_IR_tauu}
\end{align}
where we have simply substituted $\Ldecu$ in \eqref{eq:tauu_def} with \eqref{eq:Ldecu_def} and optimized over the design parameters.
If pure repetition is used, i.e., if $\ro=1$, the decoding latency and the overhead $\phi'$ are set to zero.

\subsubsection*{Decoding by the Edge Nodes}
The communication latency is given by \eqref{eq:Lde_def}. In inactivation decoding, there is a number of operations performed on the generator matrix of the code that does not scale with the number of vectors $u$ to decode. The number of decoding operations is therefore only roughly a factor $u$ larger than when the users decode and should be simulated for best precision. The decoding latency is then given by \eqref{eq:Ldece_def}. The total latency is
\begin{align}
    \taue = \min_{p, \ro, \rt} \bigg(\E[\Lc] + \frac{1}{\nce \fcpu}\E[\Nae + \Nme] + \E[\Lde]\bigg), \label{eq:rateless_IR_taue}
\end{align}
where we have simply substituted $\Ldece$ in \eqref{eq:taue_def} with \eqref{eq:Ldece_def} and optimized over the design parameters.
If pure repetition is used, i.e., if $\ro=1$, the decoding latency and the overhead $\phi'$ are set to zero.

\section{MDS Irregular-Repetition Scheme} \label{sec:MDS_IR}

Here, we propose a scheme which we refer to as the MDS irregular-repetition (MDS-IR) scheme. This scheme is similar to the Rateless-IR scheme in that it uses irregular replication, collects partial results, and has the same assignment matrix. The main difference is that the straggler code is an MDS code instead of an LT code. The MDS-IR scheme can be seen as an enhancement of the MDS-R scheme in \cite{Zhang2019,Li2021}.

Specifically, each of the length-$k$ columns of $\bs W$ is first encoded by an \([n_1, k]\) MDS code %
of  rate $\ro=k/n_1$, resulting in a matrix with rows $\rowsofC$. A coded row is then assigned to $1/\rt$ ENs on average, where $\rt=n_1/n$  is the rate of replication and $n\geq n_1$ is the total number of replicated coded rows. Note that $1/\rt$ is not necessarily an integer, as opposed to the MDS-R scheme \cite{Zhang2019,Li2021} outlined in \Cref{subsec:zhangs}. The assignment matrix is the same as for the Rateless-IR scheme in \Cref{sec:rateless_IR} and we therefore require that $k/(e\ro)$ (and $k/(e\ro\rt)$) are integers. Furthermore, recall that $\ro$ and $\rt$ are constrained by \eqref{eq:storage_constraint}. As in \Cref{sec:rateless_IR}, we can rearrange \eqref{eq:storage_constraint} and find that
\makeatletter
\if@twocolumn
\begin{align*}
    \ro &\in \bigg\{1, \frac{k}{k+1}, \frac{k}{k+2}, \ldots, \frac{k}{e\mu k}\bigg\}, \\
    \rt &\in\bigg\{1,\frac{k/\ro}{k/\ro + 1}, \ldots, \frac{k/\ro}{e\mu k}\bigg\},
\end{align*}
\else
\begin{align*}
    \ro \in \bigg\{1, \frac{k}{k+1}, \frac{k}{k+2}, \ldots, \frac{k}{e\mu k}\bigg\} \text{ and }
    \rt \in\bigg\{1,\frac{k/\ro}{k/\ro + 1}, \ldots, \frac{k/\ro}{e\mu k}\bigg\},
\end{align*}
\fi
\makeatother
where $\ro=1$ and $\rt=1$ correspond to using pure replication and pure MDS coding, respectively.

The goal is then for the ENs to compute a sufficiently large fraction of the products $\bs c_1^\T\bs X, \ldots, \bs c^\T_{n_1}\bs X$ such that $\bs w^\T_1\bs X,\ldots,\bs w^\T_k\bs X$ can be recovered. Since we are using an MDS code, any set of $k$ products is enough, but as in \Cref{sec:rateless_IR}, we note that the total latency may decrease by computing more products. This stems from two observations. Similar to the Rateless-IR scheme, computing more products may increase the spatial diversity and therefore decrease the communication latency. Furthermore, we may also recover more distinct products, which under BM decoding will decrease the decoding latency. This is because the number of additions \eqref{eq:MDS_Na} and multiplications \eqref{eq:MDS_Nm} in decoding are monotonically increasing in $F$, which is the fraction of all distinct products that were not computed. Thus, there is a trade-off between computation latency on the one hand and communication and decoding latency on the other.  To find the optimal trade-off, we use the same parameterization of the stopping set as for the Rateless-IR scheme. Specifically, we introduce a parameter $p$ and define the stopping set as the set of stopping vectors that guarantee computation of at least $k$ distinct products and at least $p$ products in total.

\subsection{Latency Analysis}
The computation latency is given by \eqref{eq:Lc_def}. We give next the communication and decoding latency of the scheme, both when the users decode and when the ENs decode. %

\subsubsection{Decoding by the Users}
The communication latency is given by \eqref{eq:Ldu_def}. From \eqref{eq:MDS_Na} and \eqref{eq:MDS_Nm}, the number of decoding operations is \(\Nau=\Na(k,\ro, F)\) and \(\Nmu=\Nm(k, \ro, F)\). The decoding latency is given by \eqref{eq:Ldecu_def}. The total latency is then given by
\makeatletter
\if@twocolumn
    \begin{align}
        \begin{split}
            \tauu = &\min_{p, \ro, \rt} \bigg(\E[\Lc] + \frac{1}{\ncu \fcpu}\E[\Na(k, \ro, F) \\
                   &\qquad\qquad\qquad\qquad  + \Nm(k, \ro, F)] + \E[\Ldu]\bigg), \label{eq:MDS_IR_tauu}
        \end{split}
    \end{align}
\else
    \begin{align}
        \tauu = \min_{p, \ro, \rt} \bigg(\E[\Lc] + \frac{1}{\ncu \fcpu}\E[\Na(k, \ro, F) + \Nm(k, \ro, F)] + \E[\Ldu]\bigg), \label{eq:MDS_IR_tauu}
    \end{align}
\fi
\makeatother
where we have simply substituted $\Ldecu$ in \eqref{eq:tauu_def} with \eqref{eq:Ldecu_def} and optimized over the design parameters.
Note that the decoding latency is zero if pure replication is used, i.e., if $\ro=1$. %

\subsubsection{Decoding by the Edge Nodes}
Each of the $M$ quickest ENs decodes $u$ vectors, such that $\Nae=u\Na(k, \ro, F)$ and $\Nme=u\Nm(k, \ro, F)$. The decoding latency is given by \eqref{eq:Ldece_def}. The communication latency is given by \eqref{eq:Lde_def} and the total latency is given by
\makeatletter
\if@twocolumn
    \begin{align}
        \begin{split}
            \taue = &\min_{p, \ro, \rt} \bigg(\E[\Lc] + \frac{u}{\nce \fcpu}\E[\Na(k, \ro, F)\\
            &\qquad\qquad\qquad\qquad + \Nm(k, \ro, F)] + \E[\Lde]\bigg), \label{eq:MDS_IR_taue}
        \end{split}
    \end{align}
\else
    \begin{align}
            \taue = \min_{p, \ro, \rt} \bigg(\E[\Lc] + \frac{u}{\nce \fcpu}\E[\Na(k, \ro, F) + \Nm(k, \ro, F)] + \E[\Lde]\bigg), \label{eq:MDS_IR_taue}
    \end{align}
\fi
\makeatother
where we have simply substituted $\Ldece$ in \eqref{eq:taue_def} with \eqref{eq:Ldece_def} and optimized over the design parameters.
Again, the decoding latency is zero if pure replication is used.  %

\section{Lower Bound on the Total Latency} \label{sec:lower_bound}
In this section, we derive a converse (lower) bound on the computation latency $\Lc$, given by \eqref{eq:Lc_def}, and on the communication latency $ \Ld$, given by \eqref{eq:Ldu_def} if the users decode, and by \eqref{eq:Lde_def} if the ENs decode. These bounds are then used to lower bound the total latency in \eqref{eq:tauu_def} and \eqref{eq:taue_def}. Note that the knowledge of the coding scheme $\calc$ and straggling times $\Lambold$ determines the computation, communication, and decoding latencies. We will therefore highlight this dependency when necessary.

We are first interested in finding a lower bound on the computation latency. To this end, consider an arbitrary coding scheme $\calc$ and straggling times $\Lambold$ resulting in $P$ computed products at the end of the computation phase. Using \eqref{eq:Di_def} and \eqref{eq:P_def} we can then express the computation latency $\Lc(\calc, \Lambold)$ as
\makeatletter
\if@twocolumn
    \begin{align}
        &\Lc(\calc, \Lambold) = \nonumber\\
        &\min \left \{t: 
        \sum_{j=1}^e \min\bigg\{\bigg\lfloor\frac{(t-\Lambda_j)^+}{\delta}\bigg\rfloor, \frac{k}{e\ro\rt} \bigg\} = P\right\}. \label{eq:Lc_conv_def}
    \end{align}
\else
    \begin{align}
           &\Lc(\calc, \Lambold) = \min \left \{t: \sum_{j=1}^e \min\bigg\{\bigg\lfloor\frac{(t-\Lambda_j)^+}{\delta}\bigg\rfloor, \frac{k}{e\ro\rt} \bigg\} = P\right\}. \label{eq:Lc_conv_def}
    \end{align}
\fi
\makeatother

Note that in \eqref{eq:Lc_conv_def}, the knowledge of the entire coding scheme is not necessary to determine $\Lc(\calc, \Lambold)$\textemdash it is sufficient to know the fraction $k/(e\ro\rt)$. We want to find the value of $k/(e\ro\rt)$ that minimizes $\Lc(\calc,\Lambold)$. The following lemma states that the optimal value is $k/(e\ro\rt)=\mu k$.

\begin{lemma}[Lower bound on $\Lc$] \label{thm:Lclower}
Consider an arbitrary coding scheme $\calc$ and straggling times $\Lambold$ resulting in $P$ computed products. Then, the computation latency $\Lc(\calc, \Lambold)$ is lower-bounded as $\Lc(\calc, \Lambold) \geq \Lclower(\Lambold,P)$, where
\begin{align*}
    \Lclower(\lambold, p) &= \min \left \{ t: \sum_{j=1}^e \min\bigg\{\bigg\lfloor\frac{(t-\lambda_j)^+}{\delta}\bigg\rfloor, \mu k \bigg\} = p \right \}.
\end{align*}

\end{lemma}
\begin{IEEEproof}
Note that in \eqref{eq:Lc_conv_def}, the summand is monotonically increasing in both $t$ and $k/(e\ro\rt)$, and $k/(e\ro\rt)\in\{k/e,\ldots,\mu k\}$ by \eqref{eq:storage_constraint} and the definition of $\ro$ and $\rt$. As $t$ decreases, $k/(e\ro\rt)$ must therefore increase for the constraint in \eqref{eq:Lc_conv_def} to hold. The minimum $t$ is thus obtained when $k/(e\ro\rt)$ is at its maximum, i.e., when $k/(e\ro\rt)=\mu k$.
\end{IEEEproof}

\begin{lemma}[Lower bound on $\Ldu$] \label{thm:Ldulower}
Assume that the decoding is carried out by the users. For any coding scheme $\calc$ and $\Lambold$ resulting in $P$ computed products, the communication latency can be lower-bounded as $\Ldu(\calc, \Lambold) \geq \Ldulower(P)$, where
\begin{align*}
    \Ldulower(p) = 
    \begin{cases}
    \cfac \frac{k^2}{p}  &\emph{if}\ p/k\in\mathbb{N}, \\
    \cfac  \frac{\ceil{p/k} k - p}{\floor{p/k}} + \frac{p - k\floor{p/k}}{\ceil{p/k}}  &\emph{otherwise}.
    \end{cases}
\end{align*}
\end{lemma}

The proof is given in Appendix \ref{app:lemma_Ldu}.

\begin{theorem}[Lower bound on $\tauu$] \label{thm:tauu_lower_bound}
Assume that the decoding is carried out by the users. For any coding scheme, the total latency can be lower-bounded as
\begin{align*}
    \tauu \geq \mathbb{E}\bigg[\min_{p\in\{k,\ldots,e\mu k\}}  \Lclower(\Lambold, p) + \Ldulower(p) \bigg],
\end{align*}
where $ \Lclower$ and $\Ldulower$ are given in \cref{thm:Lclower,thm:Ldulower}, respectively.
\end{theorem}

\begin{IEEEproof}
Using \cref{thm:Lclower,thm:Ldulower} we can write
\makeatletter
\if@twocolumn
    \begin{align*}
        \tauu &= \E[\Lc(\calc, \Lambold)] + \E[\Ldecu(\calc, \Lambold)] + \E[\Ldu(\calc,\Lambold)] \\
        &\geq \E[ \Lc(\calc,\Lambold)] + \E[\Ldu(\calc,\Lambold)] \\
        &\geq \E[\Lclower(\Lambold, P(\calc, \Lambold))] + \E[\Ldulower(P(\calc,\Lambold))] \\
        &\geq \E\bigg[ \min_{p\in\{k,\ldots,e\mu k\}} \big(\Lclower(\Lambold,p) + \Ldulower(p)\big) \bigg].
    \end{align*}
\else
    \begin{align*}
        \tauu &= \E[\Lc(\calc, \Lambold)] + \E[\Ldecu(\calc, \Lambold)] + \E[\Ldu(\calc,\Lambold)]
        \geq \E[ \Lc(\calc,\Lambold)] + \E[\Ldu(\calc,\Lambold)] \\
        &\geq \E[\Lclower(\Lambold, P(\calc, \Lambold))] + \E[\Ldulower(P(\calc,\Lambold))]
        \geq \E\bigg[ \min_{p\in\{k,\ldots,e\mu k\}} \big(\Lclower(\Lambold,p) + \Ldulower(p)\big) \bigg].
    \end{align*}
\fi
\makeatother
In the first inequality, $\Ldecu$ is lower-bounded by zero. The second inequality follows from \cref{thm:Lclower,thm:Ldulower}. At this point, the expression is still dependent on the coding scheme $\calc$ through variable $P$, while we want a bound that holds for any coding scheme. By minimizing over $p$ in the last inequality, we therefore effectively minimize over all coding schemes.
\end{IEEEproof}

\begin{lemma}[Lower bound on $\Lde$] \label{thm:Ldelower}
Assume that the decoding is carried out by the ENs. For any coding scheme $\calc$ and $\Lambold$, $\Lde$ can be lower-bounded as
\makeatletter
\if@twocolumn
\begin{align*}
     \Lde \geq \frac{uk\log_2(q)}{\nu e}.
\end{align*}
\else
$\Lde \geq \frac{uk\log_2(q)}{\nu e}$.
\fi
\makeatother
\end{lemma}

\begin{IEEEproof}
Note that in \eqref{eq:Lde_def}, $\Lde$ decreases as $M$ increases, and $M\leq e$.
\end{IEEEproof}

\begin{theorem}[Lower bound on $\taue$] \label{thm:taue_lower_bound}
Assume that the decoding is carried out by the ENs. For any coding scheme, the total latency can be lower-bounded as
\begin{align*}
    \taue \geq \mathbb{E}[\Lclower(\Lambold, k)] + \frac{uk\log_2(q)}{\nu e},
\end{align*}
where $\Lclower(\lambold, k)$ is given in \cref{thm:Lclower}.
\end{theorem}

\begin{IEEEproof}
We have
\makeatletter
\if@twocolumn
    \begin{align*}
        \taue &\geq \E[\Lc(\calc, \Lambold)] + \E[\Ldece(\calc, \Lambold)] + \E[\Lde(\calc,\Lambold)] \\
        &\geq \mathbb{E}[ \Lclower(\Lambold, P(\calc,\Lambold))] +  \frac{uk\log_2(q)}{\nu e} \\
        &\geq \mathbb{E}[ \Lclower(\Lambold, k)] +  \frac{uk\log_2(q)}{\nu e} .
    \end{align*}
\else
    \begin{align*}
        \taue &\geq \E[\Lc(\calc, \Lambold)] + \E[\Ldece(\calc, \Lambold)] + \E[\Lde(\calc,\Lambold)]
        \geq \mathbb{E}[ \Lclower(\Lambold, P(\calc,\Lambold))] +  \frac{uk\log_2(q)}{\nu e} \\
        &\geq \mathbb{E}[ \Lclower(\Lambold, k)] +  \frac{uk\log_2(q)}{\nu e} .
    \end{align*}
\fi
\makeatother
In the second inequality we lower bound $\Ldece$ by zero and use \cref{thm:Lclower,thm:Ldelower}. Since $P\geq k$ and $\Lclower$ in \cref{thm:Lclower} is monotonically increasing in $P$, the last inequality follows.
\end{IEEEproof}

\section{Numerical Results} \label{sec:results}
In this section, we numerically evaluate the latency of the proposed schemes as a function of $k$ and the straggling parameter $\beta$. We compare the schemes to the MDS-R scheme presented in \Cref{subsec:zhangs} \cite{Zhang2019}, to the block-diagonal scheme from \cite{Kumar2021_arxiv}\textemdash which we refer to as the block-diagonal irregular-repetition (BD-IR) scheme\textemdash, as well as to the lower bound derived in \Cref{sec:lower_bound}. The latency is normalized by the time it takes a user $j$ to compute $\bs y_j=\bs W\bs x_j$ by itself, which we will refer to as the local computation latency and denote by $\psi$. Since the product $\bs W\bs x_j$ requires $k(r-1)$ additions and $kr$ multiplications over $\GFq$, we have $\psi=k(2r-1)/(\ncu \fcpu)$ seconds. The local computation latency will be used to determine if there are parameter combinations for which it is actually quicker to let the users perform the inference themselves. Hereafter, any mention of a normalized latency means that the latency in question is normalized by $\psi$.

\subsection{Latency as a Function of the Matrix Size} \label{sec:latency_versus_problem_size}
In \cref{fig:latency_vs_problem_size}, we plot the total latency of the schemes as a function of the matrix size $k$. The top figure shows the performance when decoding is done by the users and the bottom figure when decoding is done by the ENs. The parameters are $e=5$, $u=10$, $k=r$, $\mu=0.6$, $\beta=30$ milliseconds (ms), $\nu=100$ Mbit/s, $\fcpu=2.7$ GHz, $\nce=50$ cores, and $\ncu=2$ cores. For simplicity, all operations are over $\GF {2}$. Furthermore, we also assume    that  there  exist  nontrivial (i.e., non-replication and non-single parity-check) binary MDS codes (in particular, RS codes). This assumption is relevant for the MDR-R and MDS-IR schemes and may result in performance curves for these schemes that are lower bounds on the actual performance. This issue will be discussed further in \cref{subsec:design_params,sec:latency_straggling_param} below. Note also that the BD-IR scheme for $q=2$  reduces to pure repetition with a different scheduling than the MDS-IR scheme. %
For the Rateless-IR scheme, we use a binary code  (which can be decoded using only additions) and the target failure probability is set to $\Pf=10^{-5}$.  By grid search optimization, we find that the optimal parameters of the robust Soliton distribution are $\gamma=210$ and $\zeta=10^{-4}$ for all $k$ when the users decode, and $\gamma=220$ and $\zeta=10^{-2}$ for all $k$ when the ENs decode. We observe for all schemes that the computation latency dominates for all $k$, that the decoding latency is increasingly important as $k$ grows, and that the communication latency is negligible for all $k$. This holds both when the users decode and when the ENs decode. We next quantify the performance differences between the schemes and compare them to the lower bound and to the local computation latency.

\subsubsection{Performance Comparison} \label{sec:performance_eval}
In the top figure, we note that the Rateless-IR scheme introduced in Section~\ref{sec:rateless_IR} outperforms the other schemes for all $k$. At $k=7000$, for example, the scheme has a $35\%$ lower latency than the MDS-R scheme \cite{Zhang2019} and a $23\%$ lower latency than the MDS-IR scheme introduced in Section~\ref{sec:MDS_IR}. The BD-IR scheme, which reduces to pure repetition for $q=2$, performs slightly worse than the MDS-IR scheme due to inferior scheduling.
As $k$ grows, the Rateless-IR scheme increases its gain compared to the MDS-R scheme, measuring at roughly $48\%$ lower latency at $k=15000$.
On the contrary, the Rateless-IR scheme has lost slightly to the MDS-IR scheme (compared to $k=7000$), measuring at only $15\%$ lower latency for $k=15000$. %
The MDS-R scheme has the worst performance for all $k$. We next compare the schemes to the lower bound and to the local computation latency.
\makeatletter
\if@twocolumn
\begin{figure}[t]
    \centering
    \begin{tikzpicture}
        \begin{axis}[%
            width = \columnwidth,
            height = 0.8\columnwidth,
            grid style = {Gray, opacity=1.0, dotted},
            xmajorgrids,
            ymajorgrids,
            xmin = 5000,
            xmax = 15000, 
            xmajorticks = false,
            xtick distance = 1000,
            ymin = 0, 
            ymax = 2.5,
            ylabel = {$\tauu/\psi$},
            legend cell align = left,
            legend style = {at={(axis cs: 15000, 2.5)}, anchor=north east, font=\scriptsize},
            ]

            \addplot[] coordinates {(5000, 1) (15000, 1)};
            \addlegendentry{Local computation}

            \addplot[MDSR] table[x=k, y=t] {./data_sweep_problem_size/SPS_MDSR_e5_mu0.6_eta0.03_ncores50_cpufreq2700000000_user.txt};
            \addlegendentry{MDS-R \cite{Zhang2019}}

            \addplot[MDSRD] table[x=k, y=t] {./data_sweep_problem_size/SPS_MDSRD_e5_mu0.6_eta0.03_ncores50_cpufreq2700000000_user.txt};
            \addlegendentry{MDS-IR}
            
            \addplot[LTR] table[x=k, y=ltot] {./data_sweep_problem_size/sweep_k_LT_REP_e5_mu0.6_eta0.03_nce50_ncu2_cpu2.7GHz_iter5e4_user.txt};
            \addlegendentry{Rateless-IR}

            \addplot[BD-IR] table[x=m, y=delay] {./data_sweep_problem_size/Sidd_results_block_diag_userdec_K5_mu0.600000_eta33.333333_phi0_u10_cores2.txt};
            \addlegendentry{BD-IR \cite{Kumar2021_arxiv}}

            \addplot[converse] table[x=k, y=t] {./data_sweep_problem_size/SPS_converse_e5_mu0.6_eta0.03_ncores50_cpufreq2700000000_user.txt};
            \addlegendentry{Lower bound}

            \node at (axis cs: 0.51e4, 2.3656) [nodetext, color=Fuchsia] {$(2, 7/10, 1/2)$};
            \node at (axis cs: 0.71e4, 1.5594) [nodetext, color=Fuchsia] {$(3, 1, 1/3)$};

            \node at (axis cs: 0.57e4, 2.21) [nodetext, color=RoyalBlue] {$(k, 1, 1/3)$};
            \addplot[] coordinates {(0.51e4, 2.23) (0.57e4, 2.21)};

            \node at (axis cs: 0.67e4, 1.85) [nodetext, color=Peach] {$(1.2k, 1/3, 1)$};
            \addplot[] coordinates {(0.51e4, 1.75) (0.68e4, 1.85)};

        \end{axis}

        \begin{axis}[%
            yshift = -0.66\columnwidth,
            width = \columnwidth,
            height = 0.8\columnwidth,
            grid style = {Gray, opacity=1.0, dotted},
            xmajorgrids,
            ymajorgrids,
            xmin = 5000,
            xmax = 15000, 
            xtick distance = 1000,
            xlabel = {$k$},
            ymin = 0, 
            ymax = 2.5,
            ylabel = {$\taue/\psi$},
            ]

            \addplot[] coordinates {(5000, 1) (15000, 1)};

            \addplot[MDSR] table[x=k, y=t] {./data_sweep_problem_size/SPS_MDSR_e5_mu0.6_eta0.03_ncores50_cpufreq2700000000_edge.txt};

            \addplot[MDSRD] table[x=k, y=t] {./data_sweep_problem_size/SPS_MDSRD_e5_mu0.6_eta0.03_ncores50_cpufreq2700000000_edge.txt};

            \addplot[LTR] table[x=k, y=ltot] {./data_sweep_problem_size/sweep_k_LT_REP_e5_mu0.6_eta0.03_nce50_ncu2_cpu2.7GHz_iter5e4_edge.txt};

            \addplot[BD-IR] table[x=m, y=delay] {./data_sweep_problem_size/Sidd_results_block_diag_edgedec_K5_mu0.600000_eta33.333333_phi0_u10_cores50.txt};
            
            \addplot[converse] table[x=k, y=t] {./data_sweep_problem_size/SPS_converse_e5_mu0.6_eta0.03_ncores50_cpufreq2700000000_edge.txt};

            \node at (axis cs: 0.6e4, 2.0) [nodetext, color=Fuchsia] {$(2, 7/10, 1/2)$};
            \addplot[] coordinates {(0.51e4, 1.96) (0.6e4, 2.0)};
            \node at (axis cs: 1.15e4, 0.9) [nodetext, color=Fuchsia] {$(3, 9/10, 1/2)$};
            \addplot[] coordinates {(1.1e4, 0.76) (1.15e4, 0.9)};
            \node at (axis cs: 0.6e4, 2.3) [nodetext, color=RoyalBlue] {$(k, 1, 1/3)$};
            \addplot[] coordinates {(0.51e4, 2.21) (0.6e4, 2.3)};

            \node at (axis cs: 0.63e4, 1.80) [nodetext, color=Peach] {$(1.2k, 1/3, 1)$};
            \addplot[] coordinates {(0.51e4, 1.73) (0.64e4, 1.80)};

        \end{axis}
    \end{tikzpicture}
    \caption{Total latency as a function of $k$ for $e=5$, $u=10$, $k=r$, $\mu=0.6$, $\beta=30$ ms, $\nu=100$ Mbit/s, $\fcpu=2.7$ GHz, $ \nce=50$, and $\ncu=2$. All operations are over $\GF {2}$. For the Rateless-IR and MDS-IR schemes the triplet $(p, \ro, \rt)$ is shown in text. For the MDS-R scheme the triplet $(\xi, \ro, \rt)$ is shown. A given triplet holds for increasing $k$ until a new triplet is given. The values of $\ro$ and $\rt$ are only approximate.}
    \label{fig:latency_vs_problem_size}
\end{figure}
\fi
\makeatother

Consider still the top figure. Clearly, the Rateless-IR scheme performs closest to the bound. %
We observe that the MDS-R scheme looses performance relative to the bound as $k$ grows. Specifically, at $k=5000$ the bound has a $29\%$ lower latency than the MDS-R scheme, which grows to roughly $40\%$ in the interval $7000\leq k \leq 15000$. We also observe that the bound has roughly $24\%$ lower latency than the MDS-IR scheme for all $k$. Note that all schemes have some value of $k$ below which a local computation would be faster\textemdash this value is $k \approx 7000$ for the Rateless-IR scheme, $k \approx 8000$ for the BD-IR and MDS-IR schemes, and $k \approx 9000$ for the MDS-R scheme. This is because neither coding nor the processing power of the ENs can speed up the computation phase enough so that it compensates for the added communication and decoding latencies. Offloading can however offer a speed-up compared to local computation of around $50\%$ or more, as can be seen for all schemes at $k=15000$. This speed-up increases as $k$ grows even larger.

Consider the bottom figure. Note that the time it takes for an EN to perform an addition or multiplication  over $\GFq$ is $\frac{1}{\ncu\fcpu}/\frac{1}{\nce\fcpu}=25$ times shorter than a user, but that an EN has to decode $u=10$ more vectors. We therefore expect the decoding latency to be roughly a factor $\frac{1}{\ncu\fcpu}/\frac{u}{\nce\fcpu}=2.5$ lower in the bottom figure compared to the top figure. Our results indeed show this, but it is hard to observe  in the figure except for the MDS-R scheme. This is because the MDS-R scheme is the only scheme with a significant decoding latency; the decoding latency of the Rateless-IR scheme is negligible and is zero for the BD-IR and MDS-IR schemes (since they  degenerate into using pure repetition, as discussed above and in \Cref{subsec:design_params} below, respectively). 

There is one change to note, however. The change being that for $k=5000$ and $k=6000$ the MDS-R scheme outperforms the MDS-IR scheme. As in the top figure, we note that the MDS-IR scheme slightly outperforms the BD-IR scheme for all $k$, which is due to inferior scheduling. The Rateless-IR scheme performs best for all values of $k$. %
Furthermore, as for the case where decoding is done by the users, the Rateless-IR scheme performs very close to the lower bound for all values of $k$. We see that the $k$ below which local computation is faster has only changed for the MDS-R scheme\textemdash it is now $k \approx 8500$ as opposed to $k \approx 9000$ in the top figure. We next focus on the annotations in \Cref{fig:latency_vs_problem_size}.

\subsubsection{Optimal Design Parameters} \label{subsec:design_params}
In both the top and bottom figures, each curve is annotated by a triplet of design parameters\textemdash $(p,\ro,\rt)$ for the proposed Rateless-IR and MDS-IR schemes, and $(\xi,\ro,\rt)$ for the MDS-R scheme. The design parameters are the result of the optimizations in \eqref{eq:rateless_IR_tauu} and \eqref{eq:rateless_IR_taue} for the Rateless-IR scheme, \eqref{eq:MDS_IR_tauu} and \eqref{eq:MDS_IR_taue} for the MDS-IR scheme, and \eqref{eq:MDS_R_tauu} and \eqref{eq:MDS_R_taue} for the MDS-R scheme. A given triplet holds for increasing $k$ until a new triplet is given. Note that for the Rateless-IR and MDS-IR schemes, the parameter $p$ is parameterized by $k$. Note also that the given rates $\ro$ and $\rt$ are rounded to simple fractions for readability. We focus next on the optimal design parameters of each scheme in turn. To this end, recall that the computation latency is the most significant. We therefore expect optimal design parameters that result in collecting products from as few ENs as possible. Due to \eqref{eq:storage_constraint}, one EN can store at most $\mu k=0.6k$ coded rows, but at least $k$ products need to be collected (exactly $k$ products if an MDS code is used). This means that products from at least two ENs need to be collected.

Consider first the Rateless-IR scheme. In both the top and bottom figure, we see that the optimal parameters are $p=1.2k$, $\ro=\frac{1}{3}$, and $\rt=1$, for all $k$. Recall that the scheme waits for $k+\phi'$ distinct products and $p$ products in total to complete, where $k+\phi'=2\mu k$. Since $\mu=0.6$, we have $k+\phi'=1.2k$. Note also that since $\ro<1$ and $\rt=1$, the scheme uses pure LT coding and no replication, which means that all computed products are distinct. As such, the scheme simply waits for the first $1.2k$ products to complete. Furthermore, note that two ENs store $\frac{2k}{e\ro\rt} = 1.2k$ products. The scheme thus has the ability to stop computing after two ENs have completed their products (although it may collect products from other ENs as well, this depends on the straggling times). The computation latency is therefore very low.

On the other hand, the MDS-IR scheme has optimal parameters $p=k$, $\ro=1$, and $\rt=\frac{1}{3}$ for all $k$, in both the top and bottom figures. This means that the scheme uses only replication and no MDS coding, and thus the assumption of the existence of nontrivial binary MDS codes (in particular, RS codes) is not  relevant. The reason for this is that the latency of decoding the MDS code is too high.

Consider now the MDS-R scheme. In the top figure for small $k$, the optimal parameters are $\xi=2$, $\ro=\frac{7}{10}$, and $\rt=\frac{1}{2}$. The scheme thus waits for two ENs and uses a combination of MDS coding and replication. As opposed to the MDS-IR scheme, there is no design parameter combination that includes both waiting for two ENs and using pure replication in order to have zero decoding latency. This is because at least one of the restrictions on the parameters presented in \Cref{subsec:zhangs} and \Cref{app:zhangs} is not met in that case. The scheme can be seen as trying to circumvent this problem by choosing a high rate for the MDS code. At $k=7000$, the scheme switches to pure replication, but then has to wait for three ENs instead. In the bottom figure, the optimal parameters are $\xi=2$, $\ro=\frac{7}{10}$, and $\rt=\frac{1}{2}$ for small and medium values of $k$, but $\xi=3$, $\ro=\frac{9}{10}$, and $\rt=\frac{1}{2}$ for $k\geq 11000$. The switch occurs in order to lower the decoding latency, which becomes more important as $k$ grows. Lastly, we remark that since the outer code rates in the top figure for $k=5000$ and $k=6000$ and for all $k$ in the bottom figure require the existence of nontrivial binary MDS codes (in particular, RS codes), these parts of the performance  curves for the MDS-R scheme are not achievable and constitute only a lower bound.
\makeatletter
\if@twocolumn
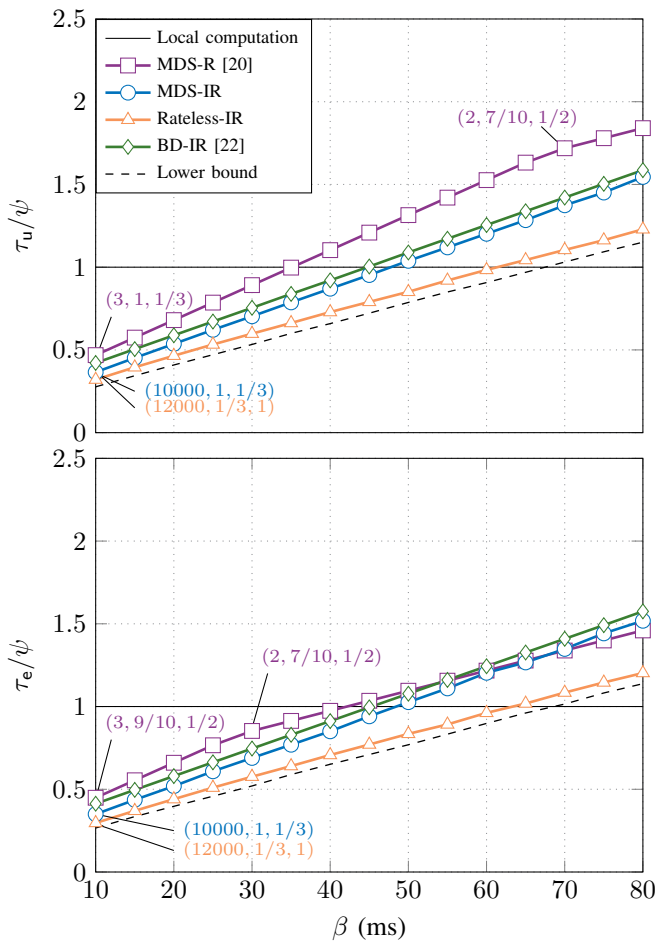
\begin{figure}[t]
\centering
\begin{tikzpicture}
\begin{axis}[%
width = \columnwidth,
height = 0.8\columnwidth,
grid style = {Gray, opacity=1.0, dotted},
xmajorgrids,
ymajorgrids,
xmin = 10,
xmax = 80, 
xmajorticks = false,
ymin = 0, 
ymax = 2.5,
ylabel = {$\tauu/\psi$},
legend cell align = left,
legend style = {at={(axis cs: 10, 2.5)}, anchor=north west, font=\scriptsize},
]

\addplot[] coordinates {(10, 1) (80, 1)};
\addlegendentry{Local computation}

\addplot[MDSR] table[x=eta, y=t] {./data_sweep_straggling_parameter/SSP_MDSR_e5_k10000_mu0.6_ncores50_cpufreq2700000000_user.txt};
\addlegendentry{MDS-R \cite{Zhang2019}}

\addplot[MDSRD] table[x=eta, y=t] {./data_sweep_straggling_parameter/SSP_MDSRD_e5_k10000_mu0.6_ncores50_cpufreq2700000000_user.txt};
\addlegendentry{MDS-IR}

\addplot[LTR] table[x=beta, y=ltot] {./data_sweep_straggling_parameter/sweep_beta_LT_REP_e5_u10_k1e4_mu0.6_phi2000_rate1e8_cpu2.7e9_iter5e4_user.txt};
\addlegendentry{Rateless-IR}

\addplot[BD-IR] table[x=beta, y=delay] {./data_sweep_straggling_parameter/Sidd_results_block_diag_userdec_K5_m10000_mu0.600000_phi0_u10_cores2.txt};
\addlegendentry{BD-IR \cite{Kumar2021_arxiv}}

\addplot[converse] table[x=eta, y=t] {./data_sweep_straggling_parameter/SSP_converse_e5_k10000_mu0.6_ncores50_cpufreq2700000000_user.txt};
\addlegendentry{Lower bound}

\node[nodetext, color=Fuchsia] at (axis cs: 10, 0.8)  {$(3, 1, 1/3)$};
\addplot[] coordinates {(10.5, 0.51) (12, 0.7)};
\node[nodetext, color=Fuchsia] at (axis cs: 55, 1.9)  {$(2, 7/10, 1/2)$};
\addplot[] coordinates {(67, 1.85) (69, 1.76)};

\node[nodetext, color=RoyalBlue] at (axis cs: 15, 0.25)  {$(10000, 1, 1/3)$};
\addplot[] coordinates {(10.5, 0.35) (15, 0.25)};

\node[nodetext, color=Peach] at (axis cs: 15, 0.15)  {$(12000, 1/3, 1)$};
\addplot[] coordinates {(10.5, 0.35) (15, 0.15)};

\end{axis}

\begin{axis}[%
yshift = -0.66\columnwidth,
width = \columnwidth,
height = 0.8\columnwidth,
grid style = {Gray, opacity=1.0, dotted},
xmajorgrids,
ymajorgrids,
xmin = 10,
xmax = 80, 
xtick distance = 10,
xlabel = {$\beta$ (ms)},
ymin = 0, 
ymax = 2.5,
ylabel = {$\taue/\psi$},
]

\addplot[] coordinates {(10, 1) (80, 1)};

\addplot[MDSR] table[x=eta, y=t] {./data_sweep_straggling_parameter/SSP_MDSR_e5_k10000_mu0.6_ncores50_cpufreq2700000000_edge.txt};

\addplot[MDSRD] table[x=eta, y=t] {./data_sweep_straggling_parameter/SSP_MDSRD_e5_k10000_mu0.6_ncores50_cpufreq2700000000_edge.txt};

\addplot[LTR] table[x=beta, y=ltot] {./data_sweep_straggling_parameter/sweep_beta_LT_REP_e5_u10_k1e4_mu0.6_phi2000_rate1e8_cpu2.7e9_iter5e4_edge.txt};

\addplot[BD-IR] table[x=beta, y=delay] {./data_sweep_straggling_parameter/Sidd_results_block_diag_edgedec_K5_m10000_mu0.600000_phi0_u10_cores50.txt};

\addplot[converse] table[x=eta, y=t] {./data_sweep_straggling_parameter/SSP_converse_e5_k10000_mu0.6_ncores50_cpufreq2700000000_edge.txt};

\node[nodetext, color=Fuchsia] at (axis cs: 10, 0.9)  {$(3, 9/10, 1/2)$};
\addplot[] coordinates {(10, 0.44) (12, 0.8)};
\node[nodetext, color=Fuchsia] at (axis cs: 30, 1.3)  {$(2, 7/10, 1/2)$};
\addplot[] coordinates {(30, 0.87) (32, 1.2)};

\node[nodetext, color=RoyalBlue] at (axis cs: 20, 0.25)  {$(10000, 1, 1/3)$};
\addplot[] coordinates {(10, 0.35) (20, 0.25)};

\node[nodetext, color=Peach] at (axis cs: 20, 0.13)  {$(12000, 1/3, 1)$};
\addplot[] coordinates {(10, 0.29) (20, 0.13)};

\end{axis}
\end{tikzpicture}
\caption{Total latency  as a function of the straggling parameter $\beta$ for $e=5$, $u=10$, $k=r=10^4$, $\mu=0.6$, $\nu=100$ Mbit/s, $f=2.7$ GHz, $\nce=50$, and $ \ncu=2$. %
All operations are over $\GF {2}$. For the Rateless-IR and MDS-IR schemes the triplet $(p, \ro, \rt)$ is shown in text. For the MDS-R scheme the triplet $(\xi, \ro, \rt)$ is shown. A given triplet holds for increasing $\beta$ until a new triplet is given. The values of $\ro$ and $\rt$ are only approximate.}
\label{fig:latency_vs_straggling_severity}
\end{figure}
\fi
\makeatother

\subsection{Latency as a Function of the Straggling Parameter} \label{sec:latency_straggling_param}
\Cref{fig:latency_vs_straggling_severity} depicts the latency as a function of the straggling parameter $\beta$. The top figure represents decoding by the users and the bottom figure decoding by the ENs. The parameters are $e=5$, $u=10$, $k=r=10000$, $\mu=0.6$, $\nu=100$ Mbit/s, $\fcpu=2.7$ GHz, $\nce=50$ cores, and $\ncu=2$ cores. %
Again, for simplicity, all operations are over $\GF {2}$. The same analysis as for \Cref{fig:latency_vs_problem_size} holds here but flipped from left to right. This is because when $\beta$ grows, the straggling becomes more severe and the computation latency therefore increases, while the decoding latency remains the same. As for \Cref{fig:latency_vs_problem_size}, some parts of the curves for the MDS-R scheme are not achievable since the outer code rates require the existence of nontrivial binary MDS codes (in particular, RS codes). In particular, the curve for $\beta \geq 70$ ms in the top figure and the entire curve in the bottom figure are not achievable and constitute only a lower bound.

\makeatletter
\if@twocolumn
\else
\begin{figure}[ht]
    \centering
    \begin{subfigure}[ht]{0.48\columnwidth}
        \centering
    \begin{tikzpicture}
        \begin{axis}[%
            width = \columnwidth,
            height = 0.8\columnwidth,
            grid style = {Gray, opacity=1.0, dotted},
            xmajorgrids,
            ymajorgrids,
            xmin = 5000,
            xmax = 15000, 
            xmajorticks = false,
            xtick distance = 1000,
            ymin = 0, 
            ymax = 2.5,
            ylabel = {$\tauu/\psi$},
            legend cell align = left,
            legend style = {at={(axis cs: 15000, 2.5)}, anchor=north east, font=\scriptsize},
            ]
            
            \addplot[] coordinates {(5000, 1) (15000, 1)};
            \addlegendentry{Local computation}
            
            \addplot[MDSR] table[x=k, y=t] {./data_sweep_problem_size/SPS_MDSR_e5_mu0.6_eta0.03_ncores50_cpufreq2700000000_user.txt};
            \addlegendentry{MDS-R \cite{Zhang2019}}
            
            \addplot[MDSRD] table[x=k, y=t] {./data_sweep_problem_size/SPS_MDSRD_e5_mu0.6_eta0.03_ncores50_cpufreq2700000000_user.txt};
            \addlegendentry{MDS-IR}
            
            \addplot[LTR] table[x=k, y=ltot] {./data_sweep_problem_size/sweep_k_LT_REP_e5_mu0.6_eta0.03_nce50_ncu2_cpu2.7GHz_iter5e4_user.txt};
            \addlegendentry{Rateless-IR}

            \addplot[BD-IR] table[x=m, y=delay] {./data_sweep_problem_size/Sidd_results_block_diag_userdec_K5_mu0.600000_eta33.333333_phi0_u10_cores2.txt};
            \addlegendentry{BD-IR \cite{Kumar2021_arxiv}}
            
            \addplot[converse] table[x=k, y=t] {./data_sweep_problem_size/SPS_converse_e5_mu0.6_eta0.03_ncores50_cpufreq2700000000_user.txt};
            \addlegendentry{Lower bound}

            \node at (axis cs: 0.51e4, 2.3656) [nodetext, color=Fuchsia] {$(2, 7/10, 1/2)$};
            \node at (axis cs: 0.71e4, 1.5594) [nodetext, color=Fuchsia] {$(3, 1, 1/3)$};

            \node at (axis cs: 0.57e4, 2.21) [nodetext, color=RoyalBlue] {$(k, 1, 1/3)$};
            \addplot[] coordinates {(0.51e4, 2.23) (0.57e4, 2.21)};

            \node at (axis cs: 0.67e4, 1.85) [nodetext, color=Peach] {$(1.2k, 1/3, 1)$};
            \addplot[] coordinates {(0.51e4, 1.75) (0.68e4, 1.85)};
            
        \end{axis}
    
        \begin{axis}[%
            yshift = -0.66\columnwidth,
            width = \columnwidth,
            height = 0.8\columnwidth,
            grid style = {Gray, opacity=1.0, dotted},
            xmajorgrids,
            ymajorgrids,
            xmin = 5000,
            xmax = 15000, 
            xtick distance = 1000,
            xlabel = {$k$},
            ymin = 0, 
            ymax = 2.5,
            ylabel = {$\taue/\psi$},
            ]
            
            \addplot[] coordinates {(5000, 1) (15000, 1)};
            
            \addplot[MDSR] table[x=k, y=t] {./data_sweep_problem_size/SPS_MDSR_e5_mu0.6_eta0.03_ncores50_cpufreq2700000000_edge.txt};
            
            \addplot[MDSRD] table[x=k, y=t] {./data_sweep_problem_size/SPS_MDSRD_e5_mu0.6_eta0.03_ncores50_cpufreq2700000000_edge.txt};
            
            \addplot[LTR] table[x=k, y=ltot] {./data_sweep_problem_size/sweep_k_LT_REP_e5_mu0.6_eta0.03_nce50_ncu2_cpu2.7GHz_iter5e4_edge.txt};
            
            \addplot[BD-IR] table[x=m, y=delay] {./data_sweep_problem_size/Sidd_results_block_diag_edgedec_K5_mu0.600000_eta33.333333_phi0_u10_cores50.txt};
            
            \addplot[converse] table[x=k, y=t] {./data_sweep_problem_size/SPS_converse_e5_mu0.6_eta0.03_ncores50_cpufreq2700000000_edge.txt};

            \node at (axis cs: 0.6e4, 2.0) [nodetext, color=Fuchsia] {$(2, 7/10, 1/2)$};
            \addplot[] coordinates {(0.51e4, 1.96) (0.6e4, 2.0)};
            \node at (axis cs: 1.15e4, 0.9) [nodetext, color=Fuchsia] {$(3, 9/10, 1/2)$};
            \addplot[] coordinates {(1.1e4, 0.76) (1.15e4, 0.9)};

            \node at (axis cs: 0.6e4, 2.3) [nodetext, color=RoyalBlue] {$(k, 1, 1/3)$};
            \addplot[] coordinates {(0.51e4, 2.21) (0.6e4, 2.3)};

            \node at (axis cs: 0.63e4, 1.80) [nodetext, color=Peach] {$(1.2k, 1/3, 1)$};
            \addplot[] coordinates {(0.51e4, 1.73) (0.64e4, 1.80)};
        
        \end{axis}
    \end{tikzpicture}
    \caption{Total latency as a function of $k$. The model matrix is square, i.e., $k=r$.}
    \label{fig:latency_vs_problem_size}
    \end{subfigure} \quad
    \begin{subfigure}[ht]{0.48\columnwidth}
        \centering
        \begin{tikzpicture}
            \begin{axis}[%
                width = \columnwidth,
                height = 0.8\columnwidth,
                grid style = {Gray, opacity=1.0, dotted},
                xmajorgrids,
                ymajorgrids,
                xmin = 10,
                xmax = 80, 
                xmajorticks = false,
                ymin = 0, 
                ymax = 2.5,
                ylabel = {$\tauu/\psi$},
                legend cell align = left,
                legend style = {at={(axis cs: 10, 2.5)}, anchor=north west, font=\scriptsize},
                ]

                \addplot[] coordinates {(10, 1) (80, 1)};
                \addlegendentry{Local computation}

                \addplot[MDSR] table[x=eta, y=t] {./data_sweep_straggling_parameter/SSP_MDSR_e5_k10000_mu0.6_ncores50_cpufreq2700000000_user.txt};
                \addlegendentry{MDS-R \cite{Zhang2019}}

                \addplot[MDSRD] table[x=eta, y=t] {./data_sweep_straggling_parameter/SSP_MDSRD_e5_k10000_mu0.6_ncores50_cpufreq2700000000_user.txt};
                \addlegendentry{MDS-IR}
                
                \addplot[LTR] table[x=beta, y=ltot] {./data_sweep_straggling_parameter/sweep_beta_LT_REP_e5_u10_k1e4_mu0.6_phi2000_rate1e8_cpu2.7e9_iter5e4_user.txt};
                \addlegendentry{Rateless-IR}

                \addplot[BD-IR] table[x=beta, y=delay] {./data_sweep_straggling_parameter/Sidd_results_block_diag_userdec_K5_m10000_mu0.600000_phi0_u10_cores2.txt};
                \addlegendentry{BD-IR \cite{Kumar2021_arxiv}}
   
                \addplot[converse] table[x=eta, y=t] {./data_sweep_straggling_parameter/SSP_converse_e5_k10000_mu0.6_ncores50_cpufreq2700000000_user.txt};
                \addlegendentry{Lower bound}

                \node[nodetext, color=Fuchsia] at (axis cs: 8.5, 0.9)  {$(3, 1, 1/3)$};
                \addplot[] coordinates {(10.5, 0.51) (12, 0.8)};
                \node[nodetext, color=Fuchsia] at (axis cs: 55, 2.0)  {$(2, 7/10, 1/2)$};
                \addplot[] coordinates {(67, 1.9) (69, 1.76)};

                \node[nodetext, color=RoyalBlue] at (axis cs: 15, 0.25)  {$(10000, 1, 1/3)$};
                \addplot[] coordinates {(10.5, 0.35) (16, 0.25)};

                \node[nodetext, color=Peach] at (axis cs: 15, 0.1)  {$(12000, 1/3, 1)$};
                \addplot[] coordinates {(10.5, 0.3) (16, 0.1)};

            \end{axis}

            \begin{axis}[%
                yshift = -0.66\columnwidth,
                width = \columnwidth,
                height = 0.8\columnwidth,
                grid style = {Gray, opacity=1.0, dotted},
                xmajorgrids,
                ymajorgrids,
                xmin = 10,
                xmax = 80, 
                xtick distance = 10,
                xlabel = {$\beta$ (ms)},
                ymin = 0, 
                ymax = 2.5,
                ylabel = {$\taue/\psi$},
                ]

                \addplot[] coordinates {(10, 1) (80, 1)};

                \addplot[MDSR] table[x=eta, y=t] {./data_sweep_straggling_parameter/SSP_MDSR_e5_k10000_mu0.6_ncores50_cpufreq2700000000_edge.txt};

                \addplot[MDSRD] table[x=eta, y=t] {./data_sweep_straggling_parameter/SSP_MDSRD_e5_k10000_mu0.6_ncores50_cpufreq2700000000_edge.txt};

                \addplot[LTR] table[x=beta, y=ltot] {./data_sweep_straggling_parameter/sweep_beta_LT_REP_e5_u10_k1e4_mu0.6_phi2000_rate1e8_cpu2.7e9_iter5e4_edge.txt};

                \addplot[BD-IR] table[x=beta, y=delay] {./data_sweep_straggling_parameter/Sidd_results_block_diag_edgedec_K5_m10000_mu0.600000_phi0_u10_cores50.txt};
                
                \addplot[converse] table[x=eta, y=t] {./data_sweep_straggling_parameter/SSP_converse_e5_k10000_mu0.6_ncores50_cpufreq2700000000_edge.txt};

                \node[nodetext, color=Fuchsia] at (axis cs: 9, 1.15)  {$(3, 9/10, 1/2)$};
                \addplot[] coordinates {(10, 0.44) (12, 1.05)};
                \node[nodetext, color=Fuchsia] at (axis cs: 30, 1.3)  {$(2, 7/10, 1/2)$};
                \addplot[] coordinates {(30, 0.87) (32, 1.2)};

                \node[nodetext, color=RoyalBlue] at (axis cs: 20, 0.25)  {$(10000, 1, 1/3)$};
                \addplot[] coordinates {(10, 0.35) (21, 0.25)};

                \node[nodetext, color=Peach] at (axis cs: 20, 0.1)  {$(12000, 1/3, 1)$};
                \addplot[] coordinates {(10, 0.29) (21, 0.1)};

            \end{axis}
        \end{tikzpicture}
        \caption{Total latency  as a function of the straggling parameter $\beta$. The model matrix is square with \(k=r=10^4\).}
        \label{fig:latency_vs_straggling_severity}
    \end{subfigure}
    \caption{The other parameters are $q=2$, $e=5$, $u=10$, $\mu=0.6$, $\beta=30$ ms, $\nu=100$ Mbit/s, $f=2.7$ GHz, $ \nce=50$, and $ \ncu=2$.  For the Rateless-IR and MDS-IR schemes the triplet $(p, \ro, \rt)$ is shown in text. For the MDS-R scheme the triplet $(\xi, \ro, \rt)$ is shown. A given triplet holds for increasing $k$ and $\beta$ until a new triplet is given. The values of $\ro$ and $\rt$ are only approximate.}
\end{figure}
\fi
\makeatother

\section{Conclusion}

We proposed a coding scheme consisting of the concatenation
of a rateless code and an irregular-repetition code
for low-latency distributed inference at the network edge. %
The proposed scheme achieves significantly lower total latency\textemdash encompassing computation latency, communication latency, and decoding latency\textemdash  than the scheme based on MDS codes and regular repetition recently proposed by Zhang and Simeone for all
considered scenarios; the proposed scheme outperforms the scheme by Zhang and Simeone by up to $35\%$. We also show that offloading the decoding step to the edge servers further improves latency.  We further derived a converse (lower) bound on the total latency.  Remarkably, the proposed scheme based on rateless codes performs very close to the bound.

An interesting outcome of this work is that, if latency induced by the decoding is taken into consideration,
 offloading computations to edge servers is not necessarily faster than if the users performed their computations locally. Hence, while most papers ignore the impact of the decoding on the total latency, it must be taken into consideration. If the computation task is large enough or the straggling mild enough, offloading can offer up to $50\%$ speed-up or more.

\appendices
\crefalias{section}{appendix}

\section{Extension of the MDS-R Scheme} \label{app:zhangs}

Since decoding was not taken into account in \cite{Zhang2019}, the MDS-R scheme \cite{Zhang2019} outlined in \Cref{subsec:zhangs} must be extended to compare it to the proposed schemes. We first briefly present the analysis in \cite{Zhang2019,Zhang2019improved_arxiv} of the computation latency and the spatial diversity the computed products before presenting our extended analysis of the communication and decoding latencies.

It was pointed out in \cite{Zhang2019} that the expected value of the $\xi$-th shortest straggling time is $\beta (h(e) - h(e-\xi))$, where $h(i)\triangleq \sum_{j=1}^i 1/j$ is the harmonic number. Since an EN computes $k/(e\ro\rt)$ products, each requiring $\delta$ seconds, the average computation latency is thus
\makeatletter
\if@twocolumn
\begin{align*}
    \E[\Lc] = \beta(h(e) - h(e-\xi)) + \frac{k}{e\ro\rt} \delta. 
\end{align*}
\else
$\E[\Lc] = \beta(h(e) - h(e-\xi)) + \frac{k}{e\ro\rt} \delta$. 
\fi
\makeatother
Note that the second term in this expression is a correction of the second term in \cite[Eq.~(13)]{Zhang2019}.

Any subset of $\xi$ ENs contains $\binom{\xi}{m}\binom{e-\xi}{1/\rt - m}$ distinct batches of spatial diversity $m$, where $\max \big\{0, \frac{1}{\rt} + \xi - e \big\} \leq m \leq \min \big\{\frac{1}{\rt}, \xi \big\}$ \cite{Zhang2019,Zhang2019improved_arxiv}. Since each batch contains $k/\big(\ro \binom{e}{1/\rt}\big)$ coded rows, the number of products of spatial diversity $m$ at the end of the computation phase is given by
\makeatletter
\if@twocolumn
    \begin{align}
        &g(m) = \nonumber\\
        &
  \left\{
    \begin{array}{@{}l@{\;\,}l@{}}
      \frac{\binom{\xi}{m}\binom{e-\xi}{1/\rt - m} n_1}{ \binom{e}{1/\rt}} &	\text{if} \max\big\{0, \frac{1}{\rt} + \xi - e\big\} \leq m \leq \min\big\{\frac{1}{\rt}, \xi\big\}, \\
      0 &\text{otherwise}.
    \end{array}
    \right.
     \label{eq:gm_def}
    \end{align}
\else
    \begin{align}
        g(m) = 
        \begin{cases}
        \frac{\binom{\xi}{m}\binom{e-\xi}{1/\rt - m} n_1}{\binom{e}{1/\rt}} &\text{if} \max\big\{0, \frac{1}{\rt} + \xi - e\big\} \leq m \leq \min\big\{\frac{1}{\rt}, \xi\big\}, \\
        0 &\text{otherwise}.
        \end{cases} \label{eq:gm_def}
    \end{align}
\fi
\makeatother

We now present our extended analysis of the communication and decoding latencies, by first considering decoding by the users and then decoding by the ENs. The decoding is done using the BM algorithm, outlined in \Cref{subsec:rs_codes}.

\subsection{Decoding by the Users}
By \eqref{eq:Ldu_def} and \eqref{eq:gm_def}, the average communication latency is 
\begin{align}
    \E[\Ldu] = \frac{u\log_2(q)}{\nu} \sum_{m=1}^e \frac{g(m)}{m}. \label{eq:MDS_R_Ld}
\end{align}
Note that in \cite{Zhang2019}, distinct products other than the $k$ products with highest spatial diversity are discarded, since transmitting them would only increase the communication latency. However, looking at \(\Na(k, \ro, F)\) in \eqref{eq:MDS_Na} and \(\Nm(k, \ro, F)\) in \eqref{eq:MDS_Nm}, we see that the number of decoding operations (and thus also the decoding latency) are monotonically increasing in $F$, where $F$ in our setting is the fraction of all distinct products that were not computed. Discarding products would thus mean increasing the decoding latency. We have observed in our simulations that the decoding latency is in general higher than the communication latency and we therefore decide not to discard any products. Note that the communication latency in \eqref{eq:MDS_R_Ld} is therefore not exactly the same as in \cite[Eq.~(13)]{Zhang2019}.

We determine the expression for $F$ as follows. Note that $\binom{e-\xi}{1/\rt}$ distinct batches are exclusively held by the $e-\xi$ straggling ENs. The fraction of distinct batches that were not computed is therefore $\binom{e-\xi}{1/\rt} / \binom{e}{1/\rt}$. Since batches are disjoint, this is also the expression for $F$. The decoding latency is then 
\begin{align*}
    \E[\Ldecu] = \frac{1}{\ncu\fcpu}(\E[\Na(k, \ro, F)] + \E[\Nm(k, \ro, F)])
\end{align*}
and the total latency is 
\begin{align}
    \tauu = \min_{\xi, \ro, \rt} (\E[\Lc] + \E[\Ldecu] + \E[\Ldu]). \label{eq:MDS_R_tauu}
\end{align}
Note that if pure replication is used, i.e., $\ro=1$, the decoding latency is set to zero. %

\subsection{Decoding by the Edge Nodes}
An EN requires the same number of operations as a user to decode one vector. Each EN must however decode $u$ vectors. The decoding latency is therefore 
\begin{align*}
    \E[\Ldece] = \frac{u}{\nce\fcpu}(\E[\Na(k, \ro, F)] + \E[\Nm(k, \ro, F)]).
\end{align*}
Since the decoded data is held by $\xi$ ENs, by \eqref{eq:Lde_def} the communication latency is 
\begin{align*}
    \E[\Lde] = \frac{uk\log_2(q)}{\nu\xi}
\end{align*}
and the total latency is given by 
\begin{align}
    \taue = \min_{\xi, \ro, \rt} (\E[\Lc] + \E[\Ldece] + \E[\Lde]). \label{eq:MDS_R_taue}
\end{align}
Again, if pure replication is used, i.e., if $\ro=1$, the decoding latency is set to zero. %

\section{Proof of Lemma 2} \label{app:lemma_Ldu}
Consider an arbitrary coding scheme. Assume that we have a realization $\bs\lambda$ that leads to a total of $p$ computed products, as well as to a set of computed \textit{distinct} products, which we will denote by $\calv$ and which has cardinality $v$. Let $\{m_i\}$ be the set of spatial diversities of the products in $\calv$ and denote by $\bar{m}$ their average, such that
\begin{align}
    \bar{m} v = \sum_{\forall i \colon \ciTX\in\calv} m_i. \label{eq:alpha_v}
\end{align}
Under zero-forcing precoding, the communication latency in \eqref{eq:Ldu_def} is given by 
\begin{align}
    \ldu = \alpha \sum_{\forall i \colon \ciTX\in\calv} \frac{1}{m_i} \label{eq:tD},
\end{align}
where $\alpha = u\log_2(q) / \nu$ is a positive factor. Looking at (\ref{eq:tD}), we see that $\ldu$ is completely specified by the number of products $v$ in $\calv$ and their spatial diversities. Also note that the spatial diversities are constrained by (\ref{eq:alpha_v}), since $\bar{m}$ and $v$ are fixed. The aim of this proof is now to find the combination of spatial diversities $\{m_i\}$ that minimizes $\ldu$ for a fixed $\bar{m}$ and $v$, and then to optimize over $\bar{m}$ and $v$ to provide the most beneficial constraint given by \eqref{eq:alpha_v}.

To this end, we define $v_m$ as the number of products in $\calv$ of spatial diversity $m$, where $1\leq m\leq e$. We can then express $v$, $\bar{m} v$, and $\ldu$ in terms of $v_m$ as
\begin{align}
    v &= \sum_{m=1}^e v_m, \label{eq:v_func_of_vm} \\
    \bar{m} v &= \sum_{m=1}^e m v_m, \label{eq:alpha_v_func_of_vm} \\
    \ldu &= \alpha \sum_{m=1}^e \frac{v_m}{m}. \label{eq:tD_func_of_vm}
\end{align}
Consequently, determining the best distribution of spatial diversities $\{m_i\}$ is equivalent to determining the best distribution of $\{v_m\}$.

We will now first derive a bound that holds when $\bar{m}$ is an integer, and then a second bound that holds when $\bar{m}$ is not an integer. The final bound is then a combination of these two bounds.

\subsection{Bound for Integer \texorpdfstring{$\bar{m}$}{hej}}
Consider the case when $\bar{m}$ is an integer. If all products in $\calv$ have the same spatial diversity $\bar{m}$, i.e., $v_m=v$ for $m=\bar{m}$ and zero otherwise, by \eqref{eq:tD_func_of_vm} the latency is $\alpha v/\bar{m}$. Comparing this to the latency of any other choice of $\{v_m\}$, we have
\makeatletter
\if@twocolumn
    \begin{align*}
        \frac{\ldu}{\alpha} - \frac{v}{\bar{m}}&= \sum_{m=1}^e \frac{v_m}{m} - \frac{v}{\bar{m}}\\
        &\stackrel{(a)}{=} \sum_{m=1}^e \frac{v_m}{m} - \frac{1}{\bar{m}}\sum_{m=1}^e v_m \\
        &= \sum_{m=1}^e v_m \frac{\bar{m} - m}{m \bar{m}} \\
        &= \sum_{m=1}^{\bar{m}} v_m \frac{\bar{m} - m}{m \bar{m}} + \sum_{m=\bar{m}}^e v_m \frac{\bar{m} - m}{m \bar{m}} \\
        &\geq \frac{1}{\bar{m}^2} \sum_{m=1}^{\bar{m}} v_m (\bar{m} - m) + \frac{1}{\bar{m}^2}\sum_{m=\bar{m}}^e v_m (\bar{m} - m)\\
        &= \frac{1}{\bar{m}^2}\sum_{m=1}^e v_m (\bar{m} - m) \\
        &\stackrel{(b)}{=} \frac{1}{\bar{m}^2} (\bar{m} v - \bar{m} v) \\
        &= 0,
    \end{align*}
\else
    \begin{align*}
        \frac{\ldu}{\alpha} - \frac{v}{\bar{m}} &= \sum_{m=1}^e \frac{v_m}{m} - \frac{v}{\bar{m}}
        \stackrel{(a)}{=} \sum_{m=1}^e \frac{v_m}{m} - \frac{1}{\bar{m}}\sum_{m=1}^e v_m
        = \sum_{m=1}^e v_m \frac{\bar{m} - m}{m \bar{m}} \\
        &= \sum_{m=1}^{\bar{m}} v_m \frac{\bar{m} - m}{m \bar{m}} + \sum_{m=\bar{m}}^e v_m \frac{\bar{m} - m}{m \bar{m}}
        \geq \frac{1}{\bar{m}^2} \sum_{m=1}^{\bar{m}} v_m (\bar{m} - m) + \frac{1}{\bar{m}^2}\sum_{m=\bar{m}}^e v_m (\bar{m} - m)\\
        &= \frac{1}{\bar{m}^2}\sum_{m=1}^e v_m (\bar{m} - m)
        \stackrel{(b)}{=} \frac{1}{\bar{m}^2} (\bar{m} v - \bar{m} v)
        = 0,
    \end{align*}
\fi
\makeatother
where in (a) we used (\ref{eq:v_func_of_vm}), and in (b) we used (\ref{eq:v_func_of_vm}) and (\ref{eq:alpha_v_func_of_vm}). When $\bar{m}$ is an integer, the latency $\alpha v/\bar{m}$ is thus the lowest possible.

\subsection{Bound for Noninteger \texorpdfstring{$\bar{m}$}{hej}}
Consider now the case when $\bar{m}$ is not an integer. The idea of the proof in this case is to partition $\calv$ into two sets and then apply the same calculations as in the previous case to each of the subsets.

Let us partition $\calv$ into the disjoint sets $\cala$ and $\calb$, with $\carda=(\ceilm - \bar{m})v$ and $\cardb=(\bar{m} - \floorm)v$. Denote by $\bar{m}_\cala$ and $\bar{m}_\calb$ the average spatial diversity of the products in set $\cala$ and $\calb$, respectively. Note that $\bar{m}_\cala\carda$ is the number of products in $\cala$ plus all its duplicates. The same holds for $\bar{m}_\calb\cardb$. Together they must make up $\bar{m} v$ products, i.e., all products in $\calv$ plus their corresponding duplicates. Then,
\begin{align}
    \bar{m} v &= \bar{m}_\cala(\ceilm - \bar{m})v + \bar{m}_\calb(\bar{m} - \floorm)v. \label{eq:alpha_v_func_of_alpha_A_and_B}
\end{align}
Denote by $v_m^\cala$ and $v_m^\calb$ the number of products of spatial diversity $m$ in set $\cala$ and $\calb$, respectively. Then,
\begin{align*}
    \bar{m}_\cala\carda &= \sum_{\forall i \colon \ciTX\in\cala} m_i = \sum_{m=1}^e m v_m^\cala
\end{align*}
and the same holds for $\calb$. We may also express $\ldu$ as
\makeatletter
\if@twocolumn
\begin{align}
    \ldu &= \alpha\sum_{\forall i \colon \ciTX\in\calv} \frac{1}{m_i} \nonumber \\
    &= \alpha \bigg( \sum_{\forall i \colon \ciTX\in\cala} \frac{1}{m_i} + \sum_{\forall i \colon \ciTX\in\calb} \frac{1}{m_i} \bigg) \nonumber \\
    &= \alpha \bigg( \sum_{m=1}^e \frac{v_m^\cala}{m} + \sum_{m=1}^e \frac{v_m^\calb}{m} \bigg) . \label{eq:tD_func_of_vmA_vmB}
\end{align}
\else
\begin{align}
    \ldu = \alpha\sum_{\forall i \colon \ciTX\in\calv} \frac{1}{m_i} 
    = \alpha \bigg( \sum_{\forall i \colon \ciTX\in\cala} \frac{1}{m_i} + \sum_{\forall i \colon \ciTX\in\calb} \frac{1}{m_i} \bigg) 
    = \alpha \bigg( \sum_{m=1}^e \frac{v_m^\cala}{m} + \sum_{m=1}^e \frac{v_m^\calb}{m} \bigg) . \label{eq:tD_func_of_vmA_vmB}
\end{align}
\fi\makeatother

Consider the case when all products in $\cala$ have the same spatial diversity $\floorm$ and all products in $\calb$ have the same spatial diversity $\ceilm$, i.e., when $v_m^\cala=\carda$ for $m=\floorm$ and $v_m^\calb = \cardb$ for $m=\ceilm$, and zero otherwise. In this case $\bar{m}_\cala=\floorm$ and $\bar{m}_\calb=\ceilm$, so that constraint \eqref{eq:alpha_v_func_of_alpha_A_and_B} is satisfied. Then by \eqref{eq:tD_func_of_vmA_vmB} the latency is $\alpha\Big(\frac{\carda}{\floorm} + \frac{\cardb}{\ceilm}\Big)$. We would like to compare this to the latency of any other choice of $\{v_m^\cala\}$ and $\{v_m^\calb\}$. Using the same principle as before, one can show that
\makeatletter
\if@twocolumn
    \begin{align*}
        \sum_{m=1}^e \frac{v_m^\cala}{m} - \frac{\carda}{\floorm} &\geq \frac{\floorm\carda - \bar{m}_\cala\carda}{\ceilm\floorm}, \\
        \sum_{m=1}^e \frac{v_m^\calb}{m} - \frac{\cardb}{\ceilm} &\geq \frac{\ceilm\cardb - \bar{m}_\calb\cardb}{\floorm\ceilm}.
    \end{align*}
\else
    \begin{align*}
        \sum_{m=1}^e \frac{v_m^\cala}{m} - \frac{\carda}{\floorm} \geq \frac{\floorm\carda - \bar{m}_\cala\carda}{\ceilm\floorm} \quad \text{and} \quad
        \sum_{m=1}^e \frac{v_m^\calb}{m} - \frac{\cardb}{\ceilm} \geq \frac{\ceilm\cardb - \bar{m}_\calb\cardb}{\floorm\ceilm}.
    \end{align*}
\fi
\makeatother
We can then conclude that
\makeatletter
\if@twocolumn
\begin{align*}
    \frac{\ldu}{\alpha} - \frac{\carda}{\floorm} - \frac{\cardb}{\ceilm} &= \sum_{m=1}^e \frac{v_m^\cala}{m} + \sum_{m=1}^e \frac{v_m^\calb}{m} - \frac{\carda}{\floorm} - \frac{\cardb}{\ceilm} \\
    & \hspace{-0.225cm} \geq \frac{\floorm\carda - \bar{m}_\cala\carda}{\ceilm\floorm} + \frac{\ceilm\cardb - \bar{m}_\calb\cardb}{\floorm\ceilm} \\
    & \hspace{-0.225cm}= \frac{\floorm\carda + \ceilm\cardb - (\bar{m}_\cala\carda + \bar{m}_\calb\cardb)}{\ceilm\floorm} \\
    & \hspace{-0.225cm}\stackrel{(a)}{=} \frac{\bar{m} v - \bar{m} v}{\ceilm\floorm} \\
    & \hspace{-0.225cm}= 0,
\end{align*}
\else
\begin{align*}
    \frac{\ldu}{\alpha} - \frac{\carda}{\floorm} - \frac{\cardb}{\ceilm} &= \sum_{m=1}^e \frac{v_m^\cala}{m} + \sum_{m=1}^e \frac{v_m^\calb}{m} - \frac{\carda}{\floorm} - \frac{\cardb}{\ceilm}
    \geq \frac{\floorm\carda - \bar{m}_\cala\carda}{\ceilm\floorm} + \frac{\ceilm\cardb - \bar{m}_\calb\cardb}{\floorm\ceilm} \\
    &= \frac{\floorm\carda + \ceilm\cardb - (\bar{m}_\cala\carda + \bar{m}_\calb\cardb)}{\ceilm\floorm}
    \stackrel{(a)}{=} \frac{\bar{m} v - \bar{m}  v}{\ceilm\floorm}
    = 0,
\end{align*}
\fi
\makeatother
where in (a) we used \eqref{eq:alpha_v_func_of_alpha_A_and_B}. Thus, the lowest possible latency when $\bar{m}$ is not an integer is $\alpha \Big( \frac{\carda}{\floorm}+\frac{\cardb}{\ceilm} \Big) = \alpha \Big( \frac{\ceilm v - \bar{m} v}{\floorm} + \frac{\bar{m} v -\floorm v}{\ceilm} \Big)$.

\subsection{Lower Bound on \texorpdfstring{$\ldu$}{hej}}
Letting $v'=\bar{m} v$ and combining the two bounds, we define
\begin{align*}
    f(v,v') = 
    \begin{cases}
    \alpha\frac{v^2}{v'}  &\text{if}\ v'/v\in\mathbb{N}, \\
    \alpha \Big( \frac{\ceil{v'/v}v - v'}{\floor{v'/v}} + \frac{v' - v\floor{v'/v}}{\ceil{v'/v}} \Big) &\text{otherwise}.
    \end{cases}
\end{align*}
By \Cref{lemma:monoton} (see Appendix~\ref{app:C} below), $f(v,v')$ is monotonically decreasing in $v'$ and monotonically increasing in $v$. Since $v'\leq p$ and $v\geq k$ (at least $k$ distinct products need to be collected, where equality is achieved by an MDS code), we can lower bound $f(v,v')$ by $\ldulower(p)$, where
\begin{align*}
    \ldulower(p) = 
    \begin{cases}
    \cfac \frac{k^2}{p}  &\text{if}\ p/k\in\mathbb{N}, \\
    \cfac \Big( \frac{\lceil p/k \rceil k -p}{\lfloor  p/k \rfloor} + \frac{p - k\lfloor p/k \rfloor}{\lceil p/k \rceil} \Big) &\text{otherwise}.
    \end{cases}
\end{align*}

This bound holds for any coding scheme and $\bs\lambda$ resulting in $p$ computed products.

\section{Lemma 4 With Proof} \label{app:C}
\begin{lemma} \label{lemma:monoton}
Let $a$ and $b$ be positive, real-valued variables satisfying $a\geq b$. Then, the function
\begin{align}
f(a,b) = 
    \begin{cases}
    \frac{b^2}{a} &\ \emph{if}\ a/b\in\mathbb{N}, \\
    \frac{\ceil{a/b} b-a}{\floor{a/b}} + \frac{a - b\floor{a/b}}{\ceil{a/b}} &\ \emph{otherwise}
    \end{cases} 
    \label{eq:fab_lemma}
\end{align}
is monotonically decreasing in $a$ and monotonically increasing in $b$.
\end{lemma}

\begin{IEEEproof}
Clearly, the term $b^2/a$ is monotonically decreasing in $a$ and monotonically increasing in $b$.

We now prove monotonicity in $a$ for $a/b\notin\mathbb{N}$. Let $a_1=cb$ and $a_2=(c+1)b$ for some positive integer $c$. If we only consider values of $a$ in the interval $a_1<a<a_2$ the fraction $a/b$ can not be an integer and thus $f(a,b)$ is evaluated by the second case in \eqref{eq:fab_lemma}. Furthermore, we can conclude that $\ceil{a/b} = a_2/b = c+1$ and $\floor{a/b} = a_1/b = c$. Then, 
\makeatletter
\if@twocolumn
    \begin{align}
        f(a,b) &= \frac{(c+1)b - a}{c} + \frac{a - cb}{c+1} \nonumber \\
        &= \frac{b}{c}\frac{(c+1)b-a}{b} + \frac{b}{c+1}\frac{a-cb}{b} \nonumber\\
        &= \frac{b}{c}\bigg(1-\frac{a-cb}{b}\bigg) + \frac{b}{c+1}\frac{a-cb}{b} \nonumber\\
        &= \frac{b}{c}\bigg(1-\frac{a-cb}{(c+1)b - cb}\bigg) + \frac{b}{c+1}\frac{a-cb}{(c+1)b-cb} \nonumber \\
        &= \frac{b}{c}\bigg(1-\frac{a-a_1}{a_2 - a_1}\bigg) + \frac{b}{c+1}\frac{a-a_1}{a_2-a_1} \nonumber\\
        &= \frac{b^2}{a_1}\bigg(1-\frac{a-a_1}{a_2 - a_1}\bigg) + \frac{b^2}{a_2}\frac{a-a_1}{a_2-a_1}. \label{eq:mono_a_last}
    \end{align}
\else
    \begin{align}
        f(a,b) &= \frac{(c+1)b - a}{c} + \frac{a - cb}{c+1}
        = \frac{b}{c}\frac{(c+1)b-a}{b} + \frac{b}{c+1}\frac{a-cb}{b} \nonumber \\
        &= \frac{b}{c}\bigg(1-\frac{a-cb}{b}\bigg) + \frac{b}{c+1}\frac{a-cb}{b}
        = \frac{b}{c}\bigg(1-\frac{a-cb}{(c+1)b - cb}\bigg) + \frac{b}{c+1}\frac{a-cb}{(c+1)b-cb} \nonumber\\
        &= \frac{b}{c}\bigg(1-\frac{a-a_1}{a_2 - a_1}\bigg) + \frac{b}{c+1}\frac{a-a_1}{a_2-a_1}
        = \frac{b^2}{a_1}\bigg(1-\frac{a-a_1}{a_2 - a_1}\bigg) + s\frac{b^2}{a_2}\frac{a-a_1}{a_2-a_1}. \label{eq:mono_a_last}
    \end{align}
\fi
\makeatother
Let $\bar{m} = \frac{a-a_1}{a_2-a_1}$, such that $\bar{m}$ goes from $0$ to $1$ as $a$ goes from $a_1$ to $a_2$. Then, \eqref{eq:mono_a_last} can be rewritten as
$f(a,b) = \frac{b^2}{a_1}(1-\bar{m}) + \frac{b^2}{a_2}\bar{m}$. 
This is a convex combination of the values $b^2/a_1$ and $b^2/a_2$, where $b^2/a_1 > b^2/a_2$. The function must therefore be monotonically decreasing on the interval $a_1 < a < a_2$. Now note that at the end points $(a_1,b)$ and $(a_2,b)$ of this interval, the function $f(a,b)$ is evaluated by the first case expression in \eqref{eq:fab_lemma}, yielding exactly $b^2/a_1$ and $b^2/a_2$. Thus, the first case expression fills in the points for which the discontinuous second case expression would evaluate to zero, making $f(a,b)$ continuous in $a$ on the interval $a_1\leq a\leq a_2$. Recall that $a$ takes values in $[b,\infty)$ (the set of all real numbers larger than or equal to $b$). This interval can be seen as a concatenation of intervals $[cb,(c+1)b]$ (the set of all real numbers $s$ such that $cb\leq s\leq (c+1)b$) for $c=1,2,\ldots$, and so $f(a,b)$ must be continuous and monotonically decreasing in $a$ on the entire interval $[b,\infty)$.

We now prove monotonicity in $b$. Let $d$ be a positive integer and define $b_1=a/(d+1)$ and $b_2=a/d$. Consider $b$ on the interval $b_1<b<b_2$, such that $\lceil a/b\rceil = a/b_1 = d+1$ and $\lfloor a/b\rfloor = a/b_2 = d$. Using the same principle as before one can show that 
$f(a,b) = \frac{b_1^2}{a}(1-\beta) + \frac{b_2^2}{a}\beta$, 
where $\beta$ goes from $0$ to $1$ as $b$ goes from $b_1$ to $b_2$. This is a convex combination of the values $b_1^2/a$ and $b_2^2/a$, where $b_1^2/a < b_2^2/a$. The function must therefore be increasing on the interval $b_1<b<b_2$. Again, note that at the end points $(a,b_1)$ and $(a,b_2)$ of this interval, the function $f(a,b)$ is evaluated by the first case expression in \eqref{eq:fab_lemma}, yielding exactly $b_1^2/a$ and $b_2^2/a$, respectively. As such, the first case expression fills in the points for which the discontinuous second case expression would be evaluated to zero, making $f(a,b)$ continuous in $b$ on the interval $b_1\leq b\leq b_2$. Recall that $b$ takes values in $(0, a]$. This interval can be seen as a concatenation of the intervals $\big[\frac{a}{d+1}, \frac{a}{d}\big]$, $d=1,2,\ldots$, and so $f(a,b)$ must be continuous and monotonically increasing in $b$ on the entire interval $(0,a]$.
\end{IEEEproof}

\balance 

% Generated by IEEEtran.bst, version: 1.12 (2007/01/11)

\end{document}